\documentclass[pre,
 reprint,
 amsmath,amssymb,
 aps,twocolumn, showkeys,
]{revtex4}
\usepackage{wasysym}
\usepackage{graphicx}
\usepackage{dcolumn}
\usepackage{bm}
\usepackage{epsfig,subfigure,float,pseudocode,multirow,footmisc,rotating}
\usepackage[nottoc, notlof, notlot]{tocbibind}
\usepackage{mathtools}
\usepackage{hyperref}

\begin{document}

\title{Zebrafish collective behaviour in heterogeneous environment modeled by a stochastic model based on visual perception}

\author{Bertrand Collignon}
 \email{bertrand.collignon@univ-paris-diderot.fr}
\affiliation{
 Univ Paris Diderot, Sorbonne Paris Cit\'e\\
 LIED, UMR 8236, 75013, Paris, France
}

\author{Axel S\'eguret}
\affiliation{
 Univ Paris Diderot, Sorbonne Paris Cit\'e\\
 LIED, UMR 8236, 75013, Paris, France
}

\author{Jos\'e Halloy}
 \email{jose.halloy@univ-paris-diderot.fr}
\affiliation{
 Univ Paris Diderot, Sorbonne Paris Cit\'e\\
 LIED, UMR 8236, 75013, Paris, France
}


\begin{abstract}
Collective motion is one of the most ubiquitous behaviours displayed by social organisms and has led to the development of numerous models. Recent advances in the understanding of sensory system and information processing by animals impel to revise classical assumptions made in decisional algorithms. In this context, we present a new model describing the three dimensional visual sensory system of fish that adjust their trajectory according to their perception field. Furthermore, we introduce a new stochastic process based on a probability distribution function to move in targeted directions rather than on a summation of influential vectors as it is classically assumed by most models. We show that this model can spontaneously transits from consensus to choice. In parallel, we present experimental results of zebrafish (alone or in group of 10) swimming in both homogeneous and heterogeneous environments. We use these experimental data to set the parameter values of our model and show that this perception-based approach can simulate the collective motion of species showing cohesive behaviour in heterogeneous environments. Finally, we discuss the advances of this multilayer model and its possible outcomes in biological, physical and robotic sciences.
\end{abstract}

\keywords{Collective behaviour --- Agent-based model --- Zebrafish --- Modeling}

\maketitle

\section{Introduction}

\subsection{Modeling collective motion}
Collective motion is one of the most ubiquitous collective behaviour displayed by interacting organisms such as cells \cite{FriedlAndGilmour2009, Friedletal.2004, EtienneManneville2014}, bacteria \cite{Sokolovetal.2009, BanJacob2008, Wolgemuth2008, Zhangetal.2010}, invertebrates (in locusts: \cite{Bazazietal.2008, Buhletal.2006, Simpsonetal.2006}; in ants : \cite{Deneubourgetal.1989, CouzinAndFranks2003}; in honeybees : \cite{Jansonetal.2005}) and vertebrates species (in birds: \cite{Ballerinietal.2008, LebarBajecAndHeppner2009, KingAndSumpter2012}; in fish: \cite{HemelrijkAndKunz2005, Parrishetal.2002, Beccoetal.2006}; in mammals: \cite{Fischhoffetal.2007}) including humans \cite{Helbingetal.2001, Moussaidetal.2011}. A growing interest to decipher the link between individual behaviours and collective patterns has arisen out of these numerous observations and led to the development of models simulating agents performing collective movement inspired from birds, mammals or fish, the latter being the focus of this paper. 

Different types of model exist to describe fish schooling. In self-propelled particles (SPP) models (synchronous \cite{Aoki1982, Aoki1984} or asynchronous \cite{Bodeetal.2010, Bodeetal.2011}) firstly used for computer animation \cite{Reynolds1987}, interactions between fish are mostly composed by a collision avoidance component, a alignement component and a cohesion component \cite{Couzinetal.2002, Lopezetal.2012}. Similarly, in social forces (SF) models fish are considered as Newtonian particles subjected to social and physical forces that respectively ensure the cohesion of the group and reflect the interaction (drag for example) with the environment \cite{Niwa1994, Niwa1996}. Both SPP and SF models have inspired several studies in statistical physics that aim to characterize features of a large number of interacting agents at the collective level \cite{Vicseketal.1995, Bertinetal.2006, Chateetal.2008, Chateetal.2010, Nagaietal.2015}. Finally, kinematic models describe the evolution of the trajectory of fish by a stochastic differential equation \cite{Gautraisetal.2009, Gautraisetal.2012, Zienkiewiczetal.2014, Mwaffoetal.2014}. This modelling approach has successfully described the movement of fish in closed environment and is a continuous time formulation of a particular case of random walks (RW). Random walks describe stochastic trajectories build by successive random steps that can be drawn from a uniform distribution (unbiased random walk) or a non-uniform distribution (biased random walked). These probabilistic models have a wide range of application, from simulating the brownian motion of particles to the exploratory patterns of many species including humans \cite{Moralesetal.2004, Raichlenetal.2014}.

\subsection{Information perception}

In all mentioned models of schooling, the agents move according to their conspecifics (position, speed, orientation or a combinaison). Multiple hypothesis have been proposed to calculate the subset of individuals that influence the motion of a focal fish: the metric perception includes all individuals situated within a defined distance; the topological perception includes the $n^{th}$ proximal neighbors; the perception based on Voronoi tessellation includes the fish connected to the focal fish by the Delaunay triangulation. However, while these hypothesis produce coherent movement of simulated agents, they are not sufficiently constrained by known biology. Therefore, recent works are now based on visual perception \cite{Lemassonetal.2009, Lemassonetal.2013, Strandburg-Peshkinetal.2013}. In these models, the focal fish does not interact with its neighbors according to their Cartesian coordinates but according to their representation in its visual field.

Thus, theoretical and experimental studies highlighted that the visual sensory system of fish is determinant for information transfer during collective motion. Indeed, the comparison of interaction mechanisms showed that a vision-based model outperforms others mechanisms (metric, topologic, Voronoi) in explaining experimental data \cite{Strandburg-Peshkinetal.2013}. In parallel, the increasing knowledge on the visual system of fish allow to develop model based on sensory systems more coherent with the biological reality. The characterization of the vision of cyprinids reveals that they have a wide visual coverage in the horizontal and vertical planes and an acute vision in the front-dorsal region \cite{Pitaetal.2015}. To make a step towards more realistic sensory systems, we introduce in this paper a new mechanism based on a the perception of 3D stimuli in the visual field. We simulate fish-agents that perceive stimuli (congeners, environment) according to the solid angle -an analogous of the planar angle but in 3D- that they intercept in their visual field.

\subsection{Information processing}

\begin{figure}[ht]
\centering
\includegraphics[width=.5\textwidth]{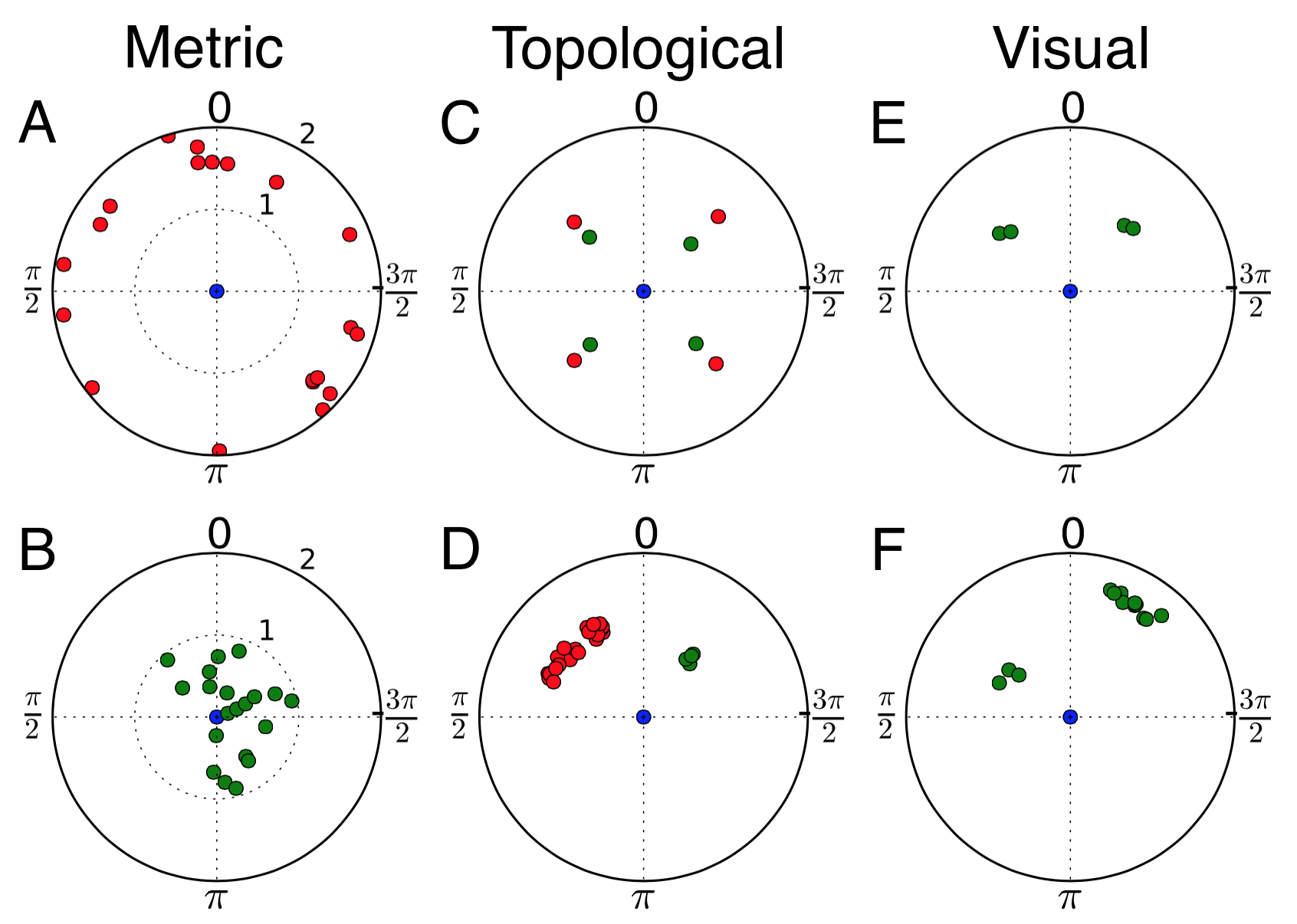}
\caption{Potential biological incoherence or problematic situations in the different models assuming a summation of interaction vectors between the blue focal fish (FF) at the center and green perceived fish (non-perceived fish are represented in red). In the metric model, the FF perceives no (A) or a high number (B) of neighbors. In the topological model, the position of the $n^{th}$ closest neighbors can be conflicting (C) or disproportionate regarding the rest of the group (D). Even in visual model, the position of the perceived fish can lead to a resulting vector oriented between the stimuli (E) or the number of fish can be overestimated by their captured angle due to proximity (F).}
\label{fig:problem}
\end{figure}

Once the focal fish has perceived potential stimuli, this information has to be processed and translated into a movement of the individual. In SPP and SF models considering only the influence of congeners, a vector of interaction is computed with all neighbors situated in a delimited range (metric models) or according to their proximity rank (topological models). Then, the focal fish moves along the resulting force computed by a weighted summation process of the different interaction forces applied on the focal fish. While it results in coordinated motion of agents, these models can also produce biological incoherent behaviour of the simulated individuals: the influence of some congeners can be omitted/overestimated or the resulting vector can point towards a direction where no stimulus is present (Fig.~\ref{fig:problem}). In addition, such process have difficulties in reproducing experimentally observed choices between two concurrent stimuli. For example, zebrafish larvae randomly chose to orient towards one of two equidistant source of light and do not follow the bisector \cite{Burgessetal.2010}. Here, we present a novel algorithm to account for information processing by the individual. Rather than summing influential vectors, we propose to sum probability distribution functions to orient towards the different stimuli. As shown in the model section, such mixture distribution can spontaneously produces transitions from choice to consensus and better describes different biological observations.

\subsection{A multilevel approach}

It would be interesting to develop a multilevel modeling approach that takes into account models for perception and information processing by the individuals. Beyond existing SPP and RW models, it requires extending the description of the agents by biologically relevant properties. We propose a new hybrid class of models, at the crossroads of both RW and SPP models, that include a probabilistic component (inspired by RW) in the behaviour of the agents that react to their perception field (inspired by SPP models). In the model, the agents chose a direction to move according to a probability distribution function (PDF) that is determined by the presence of stimuli in their visual perception field, in accordance with observed stochastic choices made by zebrafish larvae \cite{Burgessetal.2010}. This PDF is computed as a mixture distribution of von Mises distributions centered on each perceived stimulus. Once the probability distribution function has been computed, the direction of the agent is chosen accordingly to this PDF. With this model, we simulate motion of single individuals or groups of fish evolving in a bounded environment that can include other stimuli like spots of interest.

To validate this model, we measured the individual (single fish) and collective (group of 10 fish) motion of adult zebrafish in different environmental conditions. Zebrafish form loosely cohesive groups that do not show strong alignment and could be challenging to model with a classical approach. In addition, we observed the behaviour of zebrafish in a bounded tank with potential spots of interest to take into account the interactions with the environment. First, we analysed the locomotion of isolated individual evolving in a uniform environment to determine the intrinsic characteristics of their motion (speed, change in orientation and probability of presence). We performed similar measurements in heterogeneous environment by placing two floating plastic disk acting as attractive spots. Then, we investigated the influence of conspecifics on the spatial distribution of fish by observing collective motion of group of zebrafish in both homogeneous and heterogeneous environment. We measured their probability of presence in the experimental tank and their inter-individual distances. This experimental data is used for both, setting parameter values for individual behaviour, as well as to compare the predictions obtained from our model.


\section{New stochastic model}

Our aim is to model fish swimming in nearly 2D in a bounded environment with potential spots of interest that attract them. To do so, we simulate agents that update their position vector ${X_i}$ with a velocity vector ${V_i}$ though a discrete time process in a bounded two-dimensional space:

\begin{equation}
X_i(t+\delta t)= X_i(t) + V_i(t)\delta t
\label{equa:zebrapos}
\end{equation}
\begin{equation}
V_i(t+\delta t)= v_i (t+\delta t) \Theta_i (t+\delta t)
\label{equa:zebraspeed}
\end{equation}

with $v_i$ the linear speed of the $i^{th}$ agent and $\Theta_i$ its orientation. Since we focus on the decision making process of the fish to chose its orientation in a complex environment (walls of the tanks, spots of interest) with other fish, we assume the simple hypothesis that the linear speed $v_i$ of the agent is randomly drawn from the instantaneous speed distribution measured in our experiments. The novelty of this model is the computation of the orientation $\Theta_i$. We model the spherical visual perception field of fish (Fig.~\ref{figure2}) and describe the probability for fish to move in all potential directions by a circular probability density function extending from $-\pi$ to $\pi$. Thus, $\Theta_i$ is not computed as a resulting vector with potential noise but is drawn from a circular probability density function (PDF) formed by von Mises distributions, an equivalent to the gaussian distribution in circular probability. This PDF is characterized by a measure of location $\mu$ (equivalent to the mean of a Gaussian PDF) and a measure of concentration $\kappa$ (an inversely proportional equivalent to the variance of a Gaussian PDF). For a fish that perceives no stimuli, the distribution of probability is described by $\mu = 0$ while the value of $\kappa_0$ is determined experimentally (Eq.~\ref{PDFbasic}). By doing so, a fish that perceives no stimuli will move forwards with deviation inversely proportional to $\kappa_0$. Since our goal is to model fish moving in a bounded tank, we introduce the interaction of the fish with the walls in the computation of this PDF. As soon as a fish is situated in a distance shorter than a threshold value $d_w$, we assume that the fish starts a wall-following behaviour. To simulate this behaviour, the value of $\mu$ is not equal to 0 but to the direction along the followed wall. Since there are two potential direction, the PDF is computed as the weighted sum of the PDF associated with each direction. Thus, the PDF $f_{0}(\theta)$ for a fish to move in each potential direction $\theta$ in a bounded tank without perceptible stimuli is given by:

\begin{equation}
f_{0}(\theta)=
\begin{dcases}
	\frac{1}{2 \pi I_0 (\kappa_{0})} exp[\kappa_{0} cos(\theta)],& \text{if } d \geq d_w \\
	\frac{1}{2} \sum_{i=1}^{2} \frac{1}{2 \pi I_0 (\kappa_{w})} exp[\kappa_{w} cos(\theta - \mu_{w_i})],& \text{if } d < d_w \\
\end{dcases}
\label{PDFbasic}
\end{equation}

\begin{equation}
I_0(\kappa)= \sum_{k=0}^{\infty} \frac {(\frac{\kappa}{2})^{2k}}  {k! \Gamma(k+1)}
\label{besselW}
\end{equation}

with $d$ the distance with the closest wall, $d_w$ the threshold distance determining the interaction with the walls, $\kappa_{0}$ and $\kappa_{w}$ the dispertion parameters respectively associated with the basic-swimming and wall-following behaviours,  $\mu_{w_1}$ and $\mu_{w_2}$ the two possible directions along a considered wall and $I_0$ the modified Bessel function of first kind of order zero. The value of $d_w$, $\kappa_{0}$ and $\kappa_{w}$ are determined experimentally.

Equations (\ref{equa:zebrapos}) to (\ref{besselW}) are sufficient rules to simulate a fish swimming in an experimental tank. However, since we aim at simulating groups of fish moving in a homogeneous but also heterogeneous environment, we implement the interactions with other congeners and spots of interest. As soon as the fish perceive stimuli in its perception field, its PDF is influenced by those stimuli following two steps: information gathering and information processing.

\subsection{Information gathering}

\begin{figure*}[ht]
\centering
\includegraphics[width=\textwidth]{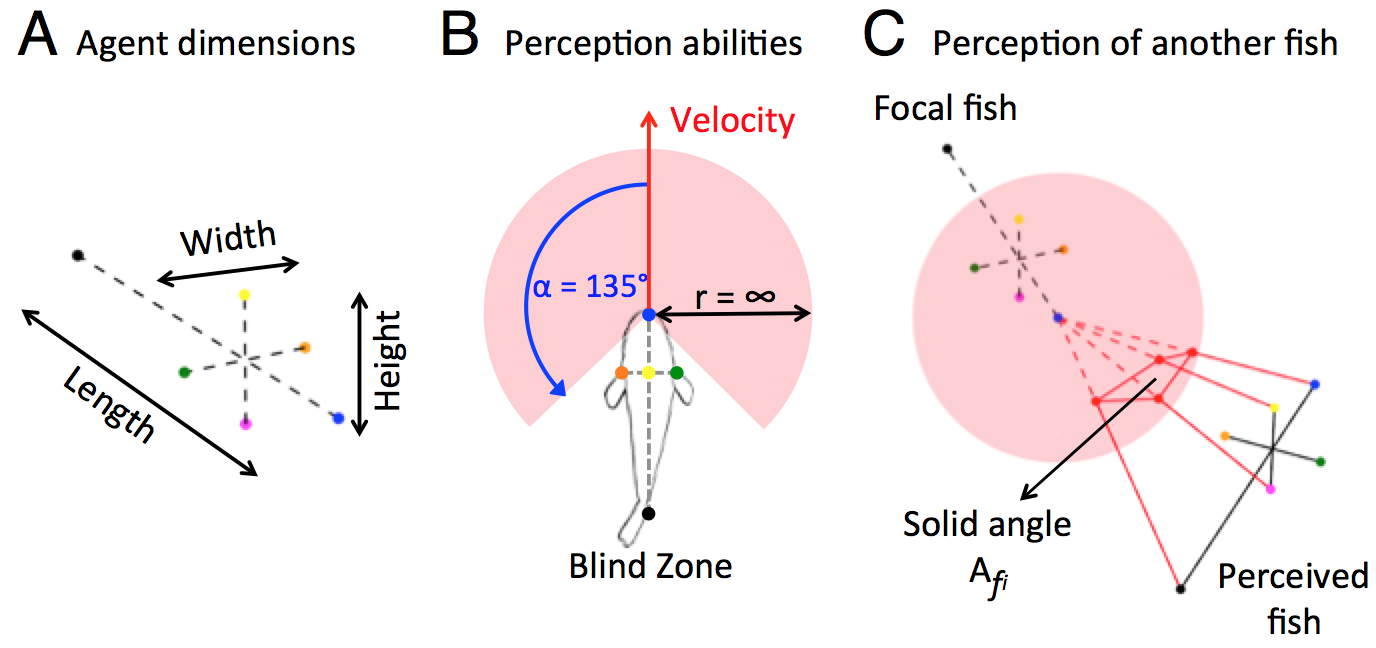}
\caption{Scheme of the simulated fish and their perception abilities. (A) Simulated fish swim on a 2D plane space and are characterized by a width (1 cm), length (3.5 cm) and height (1cm). (B) Their spherical perception field is defined by an angular perception zone and an infinite distance of perception. The position (x, y) of the fish is updated at each time step by a velocity vector computed according to the stimuli perceived. (C) Fish present in their perception field are perceived as the solid angle $A_{f_i}$ captured by their extremities. Similarly, we computed the solid angle $A_{s_i}$ captured by the spots of interest.}
\label{figure2}
\end{figure*}

We simulate fish-agents that can perceive and react to 3 dimensional stimuli perceived in their visual field. Fish are modeled as 3D polygons with 6 vertices swimming on a 2D plane space (Fig.~\ref{figure2}A). Their visual perception is simulated as a cyclopean vision sensor that has a 270 degree spherical field of view extending frontally and laterally and an infinite effective radius (Fig.~\ref{figure2}B). Fish perceive potential stimuli by the solid angle that the projection of their extremities capture in their spherical perception field (Fig.~\ref{figure2}C).

To reflect our experimental conditions, we include two potential stimuli in our simulations: fish and spots of interest. Fish are considered as polygons of length = 0.035m, width = 0.01m and height = 0.01m whose extremities form the two boundaries of the fish. Spots of interest are considered as floating disks with a radius of 0.1m floating 0.05m above the plane space.

\subsection{Information processing}

Once all potential stimuli have been perceived, the model computes a PDF for the focal fish to move according to each stimulus (fish or spots in this study). For example, the probability of the focal fish to orient towards a perceived fish is given by a von Mises distribution clustered around this fish:

\begin{equation}
f_{f_i}(\theta)= \frac{1}{2 \pi I_0 (\kappa_{f})} exp[\kappa_{f} cos(\theta - \mu_{f_i})]
\end{equation}

\begin{equation}
I_0(\kappa_{f_i})= \sum_{k=0}^{\infty} \frac {(\frac{\kappa_{f}}{2})^{2k}}  {k! \Gamma(k+1)}
\end{equation}

with $\theta$, the potential direction of movement of the fish, $\mu_{f_i}$ the location of the perceived fish $i$ and $\kappa_{f}$ a measure of concentration. 

\begin{figure}[ht]
\centering
\includegraphics[width=.5\textwidth]{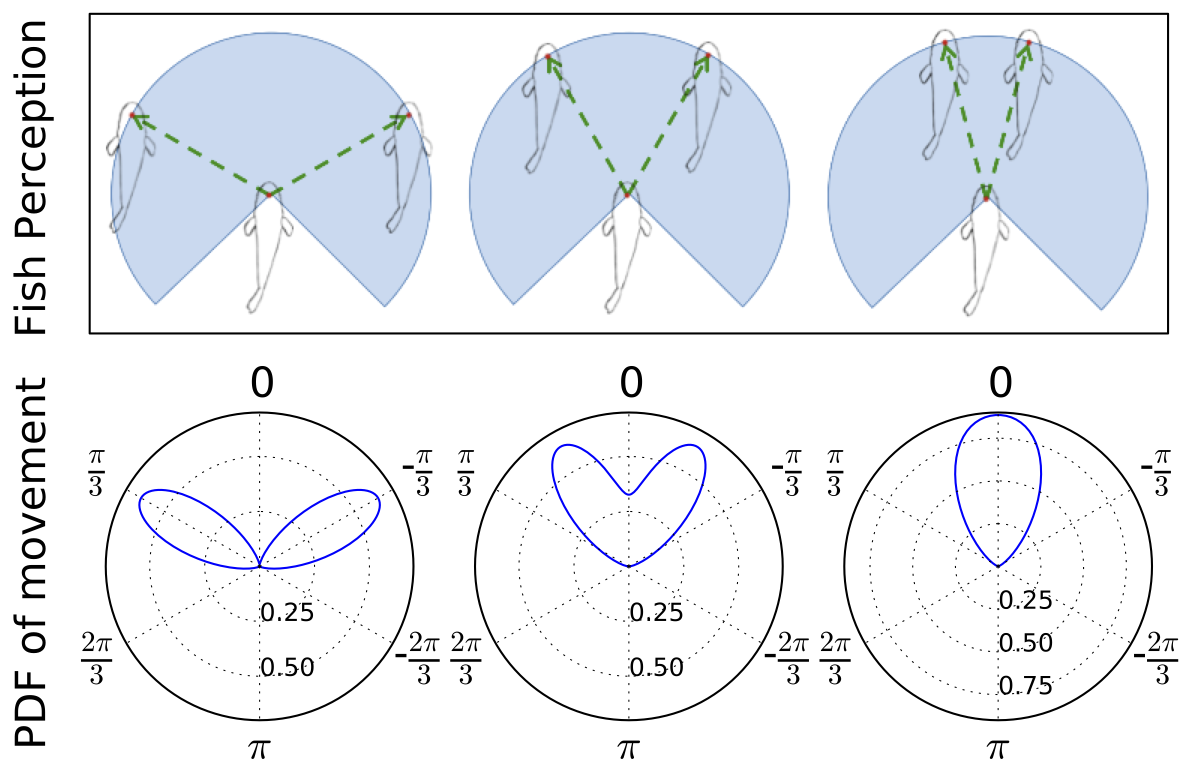}
\caption{Evolution of the total probability density function $f_F(\theta)$ according to the position of the neighboring individuals. Contrary to classical models, the fish will not always move forward as it would be predicted by weighting the attraction vectors. When the neighbors are far from each other, the focal fish will randomly orient towards one of them. On the contrary, if the two neighbors are closer, the sum of the PDF associated with to these fish predict that the focal fish will favor the direction between them rather than selecting one of them. Thus, by summing the PDFs rather than interaction vectors, the model reproduce transitions from choice to consensus and vice versa according to the angle between two perceived stimuli.}
\label{figure3}
\end{figure}

The model draws such distribution for each stimulus (fish or spot) in the perception field of the focal fish. Then for each type of stimuli, it performs a weighted sum of all distributions proportional to the ratio of the solid angle $A_{*_i}$ captures by each stimulus to the sum of the solid angles captured by all stimulus $A_{T_*}$. For example, the PDF computed for the perceived fish is given by :

\begin{equation}
f_{F}(\theta)= \sum_{i=1}^{n} \frac{A_{f_i}}{A_{T_f}} \frac{1}{2 \pi I_0 (\kappa_{f})} exp[\kappa_{f} cos(\theta - \mu_{f_i})]
\end{equation}

\begin{equation}
A_{T_f} = \sum_{i=1}^{n} A_{f_i}
\end{equation}

with $A_{T_f}$ to the sum of the solid angles $A_{f_i}$ captured by all fish. Thanks to the summation of PDFs rather than vectors, the model intrinsically produces transitions between consensus (i.e. the fish orients between two stimuli) and choice (i.e. the focal fish orients toward one of two stimuli) according to the angle between to stimuli (Fig.~\ref{figure3}).

We calculate a similar PDF for the spots of interest perceived by the focal fish:

\begin{equation}
f_{S}(\theta)= \sum_{i=1}^{n} \frac{A_{S_i}}{A_{T_s}} \frac{1}{2 \pi I_0 (\kappa_{s})} exp[\kappa_{s} cos(\theta - \mu_{s_i})]
\end{equation}

\begin{equation}
A_{T_s} = \sum_{i=1}^{n} A_{S_i}
\end{equation}

with $\mu_{s_i}$ the location of the center of the perceived spots, $\kappa_{s}$ the dispersion parameters of the perceived spots, $A_{s_i}$ the solid angle captured by the spot and $A_{T_s}$ the sum of the solid angles $A_{S_i}$ captured by all spots.

\begin{figure}[ht]
\centering
\includegraphics[width=.43\textwidth]{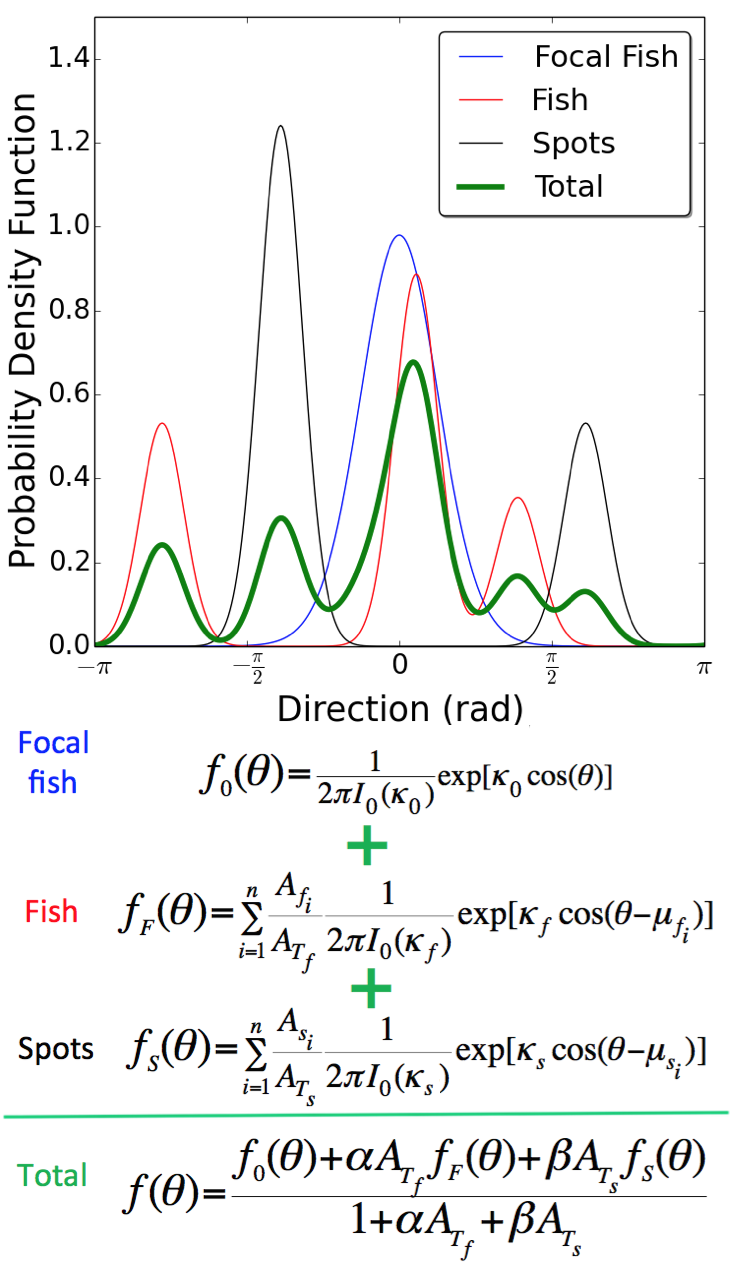}
\caption{Weighted sum of the three PDFs calculated for the focal fish according to the congeners and the shelters perceived by the individual to compute its final PDF. The direction taken by a fish is drawn randomly from the final computed PDF (in green) by inverse transform sampling.}
\label{figure4}
\end{figure}

Once the PDFs have been computed for each type of stimuli (fish - spots), we calculate a weighted sum of the PDFs to obtain the $\textit{global}$ probability distribution function $f(\theta)$ of the focal fish to move towards a given direction. In this first approach, we assumed that the weight of the PDF associated with each type of stimuli is a linear function of the total solid angle $A_{T_*}$ that they capture in the perception field of the focal fish (Fig.~\ref{figure4}). This implies that the fish respond more strongly to largely perceived stimuli but it also allows a potential hierarchy in the response to different stimuli. Thus, the $\textit{global}$ PDF f($\theta$) is given by:

\begin{equation}
f(\theta)= \frac{f_{0}(\theta) + \alpha_{*} A_{T_f} f_{F}(\theta) + \beta_{*} A_{T_s} f_{S}(\theta)} {1+ \alpha_{*} A_{T_f} + \beta_{*} A_{T_s}}
\end{equation}

\begin{equation}
\alpha_{*}=
\begin{dcases}
	\alpha_{0},& \text{if } d \geq d_w \\
	\alpha_{W},& \text{if } d < d_w \\
\end{dcases}
\end{equation}

\begin{equation}
\beta_{*}=
\begin{dcases}
	\beta_{0},& \text{if } d \geq d_w \\
	\beta_{W},& \text{if } d < d_w \\
\end{dcases}
\end{equation}

with $\alpha_{0}$ and $\beta_{0}$ the parameters weighting the influence of respectively the fish and the shelters for a fish distant from any wall and $\alpha_{W}$ and $\beta_{W}$ for a fish following a wall. These parameters are fitted experimentally. 

Then, we numerically compute the cumulated distribution function (CDF) corresponding to this custom PDF $f(\theta)$ by performing a cumulative trapezoidal numerical integration of the PDF in the interval [$-\pi$,$\pi$]. Finally, the model draws a random direction $\Theta_i$ in this distribution by inverse transform sampling. The position of the fish is then updated according to this direction and his velocity with equations~\ref{equa:zebrapos} and ~\ref{equa:zebraspeed}.


\section{Results}

\subsection{Homogeneous environment}

\begin{figure}[ht]
\centering
\includegraphics[width=.5\textwidth]{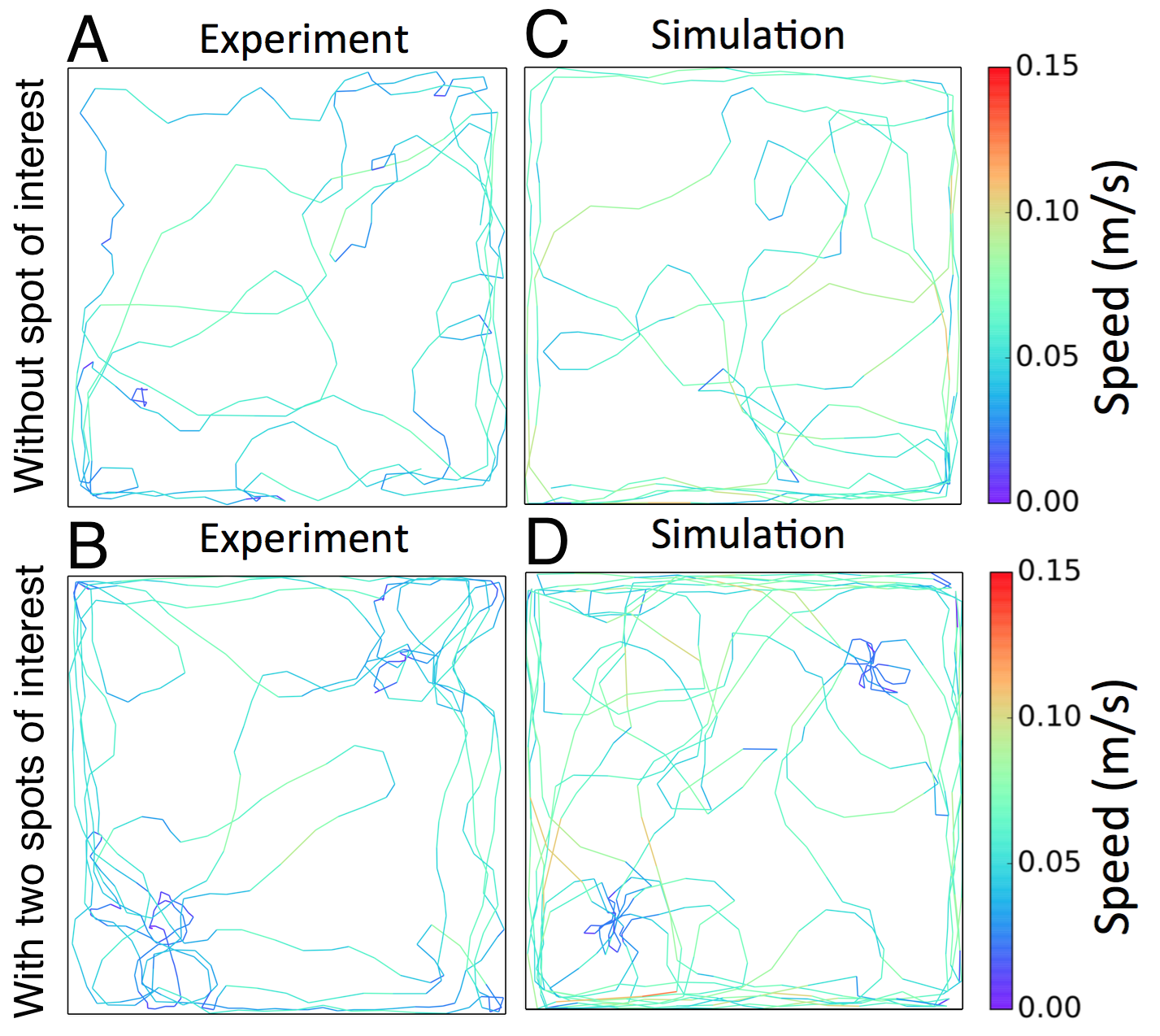}
\caption{Example of experimental (A-B) and simulated (C-D) trajectories of a fish swimming alone during 10 minutes in the absence (A-C) or presence (B-D) of two floating disks. The colour of the trajectory indicates the speed of the individual.}
\label{figure5}
\end{figure}

\begin{figure}[ht]
\centering
    {
      \includegraphics[width=0.47\textwidth]{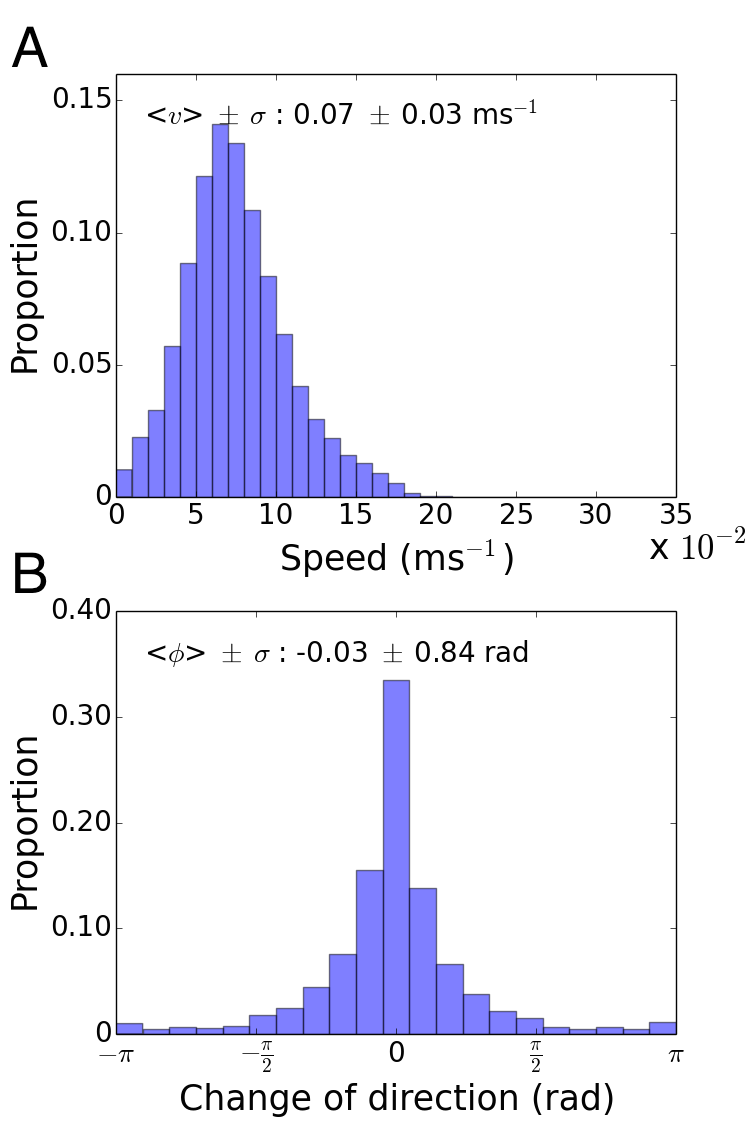}
    }
  \caption{Experimental individual behaviour of single fish from AB strain in the experimental tank without stimuli. (A) The distribution of the speed shows an average speed of 0.07 $\pm$ 0.03 ms$^{-1}$. (B) The distribution of the change in orientation highlights that fish are mainly swimming forward with low deviation. Results are cumulated for 10 replicates of one hour.}
  \label{figure6}
\end{figure}

We measured the positions of ten fish tested individually swimming alone in our 1.20m x 1.20m experimental tank during 1 hour. Based on this tracking, we build the trajectories of the fish and computed their speed and change in orientation. An example of a 10 minutes trajectory of a fish is given on Fig.~\ref{figure5}A. Fig.~\ref{figure6}A shows the cumulated distribution of all instantaneous speeds measured in a homogeneous environment with a average speed of 0.07 $\pm$ 003 ms$^{-1}$. Similarly, Fig.~\ref{figure6}B represents the cumulative distribution of the changes of orientation along the trajectories and highlights that fish are mainly moving forwards with soft change of orientation (average change = -0.03 $\pm$ 0.84 rad). The distribution of the positions detected in the tank (Fig.~\ref{figure11}A) shows that the higher probability of presence was found along the walls. Thus, in a homogeneous environment, fish were mainly swimming along the walls of the tank and avoid the centre of it. 

\begin{figure}[h!]
\centering
    {
      \includegraphics[width=0.45\textwidth]{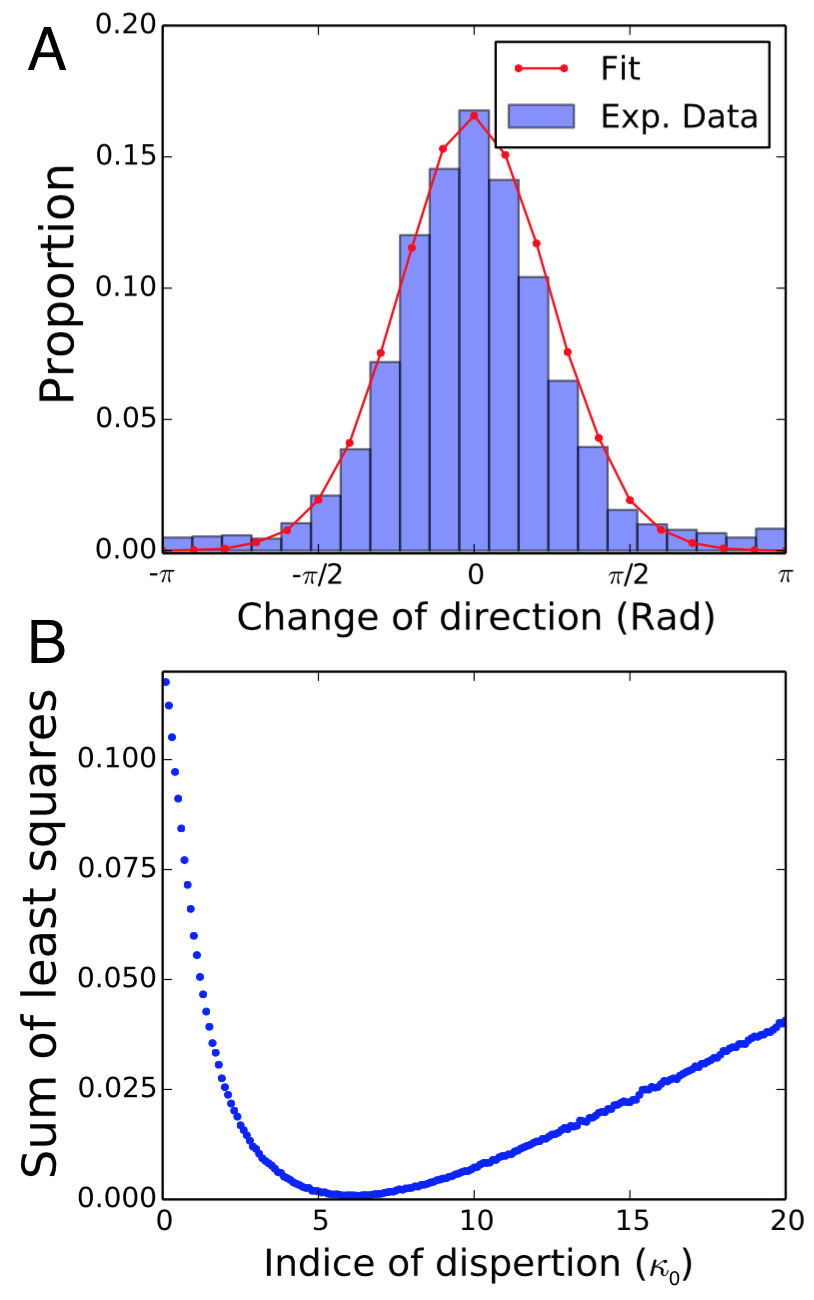}
    }
  \caption{Experimental distribution and fitting by sum of least squares minimisation of direction changes of single zebrafish swimming in a homogeneous tank. (A) Experimental distribution (blue) and fitting (red) of changes of direction obtained for $\kappa_0=6.3$. The data were calculated for trajectories measured at least at 0.3m from any walls to minimize their influence on the fish orientation. (B) Evolution of the sum of the mean squares between the experimental distribution and a theoretical distribution of 100,000 random draws in a von Mises distribution with parameters $\mu = 0$, $\kappa_0$.}
  \label{figure7}
\end{figure}

We used this experimental data to set the parameters of our model in order to simulate the movement of a single fish in a homogeneous tank. In the simulations, the speed of the agent is drawn from the experimental distribution of the instantaneous speed of the fish. Speeds are drawn independently from each other so that there is no correlation between the speed of agent $i$ at time $t$ and $t+1$. While this differs obviously from the reality, for simplicity we do not take into account speed matching in this first parametrization of our new model.
Experiments with single fish allow us to fit the parameter value characterizing the change of direction of fish $\kappa_0$. To do so, we measured the change of direction of the fish when they were at least at 0.30m from any walls. By doing so, we excluded the potential influence of the walls and consider only the intrinsic change of direction. Then, we compared this experimental distributions to theoretical ones and select the best fitting. In this first approach, we chose the sum of least squares minimization as a rapid and low computational costly fitting method. The best fitting of this experimental distribution was obtained by $\kappa_0 = 6.3$, as shown by Fig.~\ref{figure7}.
 
\begin{figure}[h!]
\centering
\includegraphics[width=.5\textwidth]{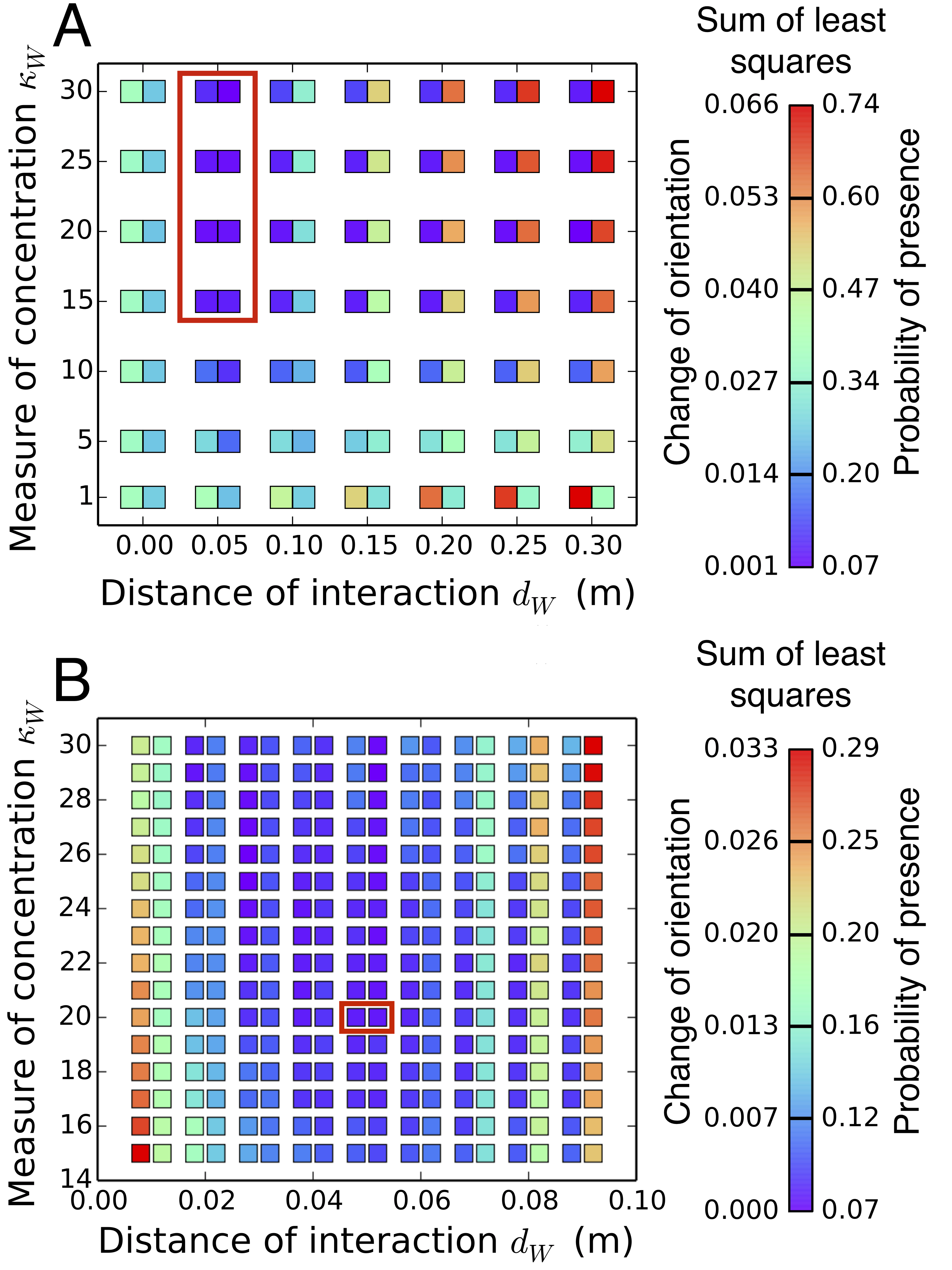}
\caption{Fitting by sum of least squares minimisation of the parameters $d_w$ and $\kappa_W$ determining the interaction of a single fish with the walls of the tank. The parameter $d_w$ corresponds to the threshold distance of interaction with the walls and $\kappa_W$ is inversely proportional to the width of the von Mises distribution associated with wall following. To each couple ($d_w$, $\kappa_W$) corresponds a couple of squares whose colours indicate the value of the sum of least squares obtained for the comparison of the changes of direction (left square) and the probability of presence (right square) with the experimental data shown in Fig.~\ref{figure6}B and Fig.~\ref{figure11}A. Lower values of sum of least squares corresponding to better fits are in the purple and blue colour scale. (A) The exploration of a first set of parameter values indicated that the best couple ($d_w$ and $\kappa_W$) are found for values ($d_w = 0.05$ and $ 15 < \kappa_W < 30$), highlighted in the red rectangle. (B) The refinement of the parameters exploration in the ranges determined in (A) showed that the best fitting is obtained by  $d_w = 0.05$ and $\kappa_W = 20$.}
\label{figure8}
\end{figure}

For experiments involving a single fish, only the interaction with the walls of the tank is present. Therefore, the other relevant parameters are the distance of interaction with the walls $d_w$ and the measure of concentration $\kappa_w$ of the PDFs associated with wall following. To estimate these parameter values, we performed simulations with different couples of value ($d_w$, $\kappa_w$) and compared the experimental distributions of change in orientation and of probability of presence with those generated by the simulations. Fitting of these parameters showed that $d_w = 0.05$ and $\kappa_w = 20$ are the best values to reproduce our experimental data (Fig.~\ref{figure8}). A 10 minutes trajectory of a simulated fish with these parameter values is shown on Fig~\ref{figure5}C.

\begin{figure}[ht]
\centering
\includegraphics[width=.4\textwidth]{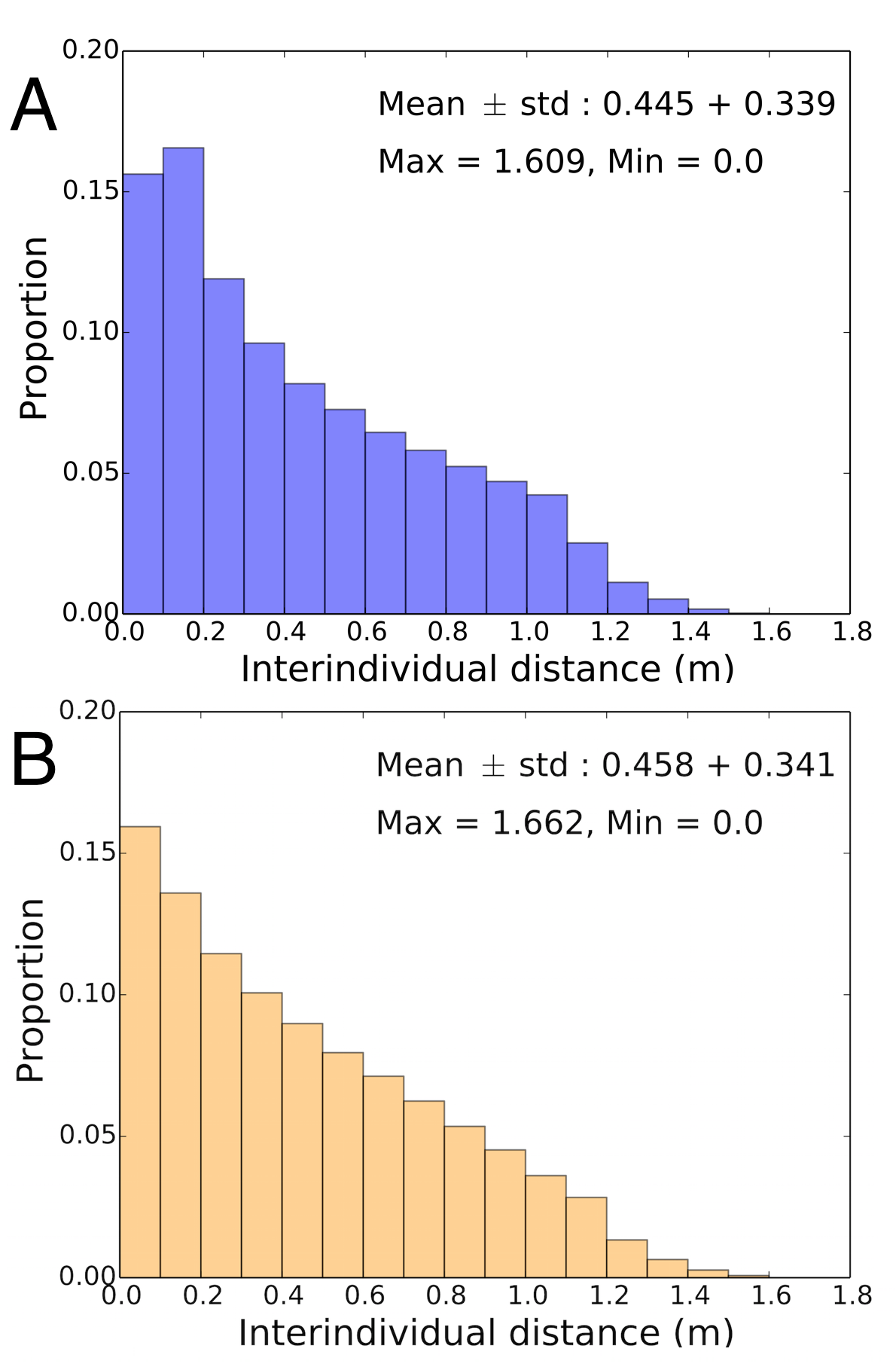}
\caption{Cumulated interindividual distances measured between all pairs of fish. Results are obtained for 10 experiments of one hour with 10 groups of 10 fish (A) and 10 simulations with 10 agents (B).}
\label{figure9}
\end{figure}

Then, we performed experiments with 10 groups of 10 zebrafish. As observed for single fish experiments, fish were mostly detected along the walls of the tank (Fig.~\ref{figure11}C). Since we did not track the fish individually when swimming in group, we could not build the individual trajectories of the fish. However, we measured the distance between all pair of fish at each time step as a measure of group cohesion. The distribution of these interindividual distances shows an average of 0.394m $\pm$ 0.38m with the mode of the distribution between 0.1 and 0.2m (Fig.~\ref{figure9}A).

\begin{figure}[h!]
\centering
\includegraphics[width=.5\textwidth]{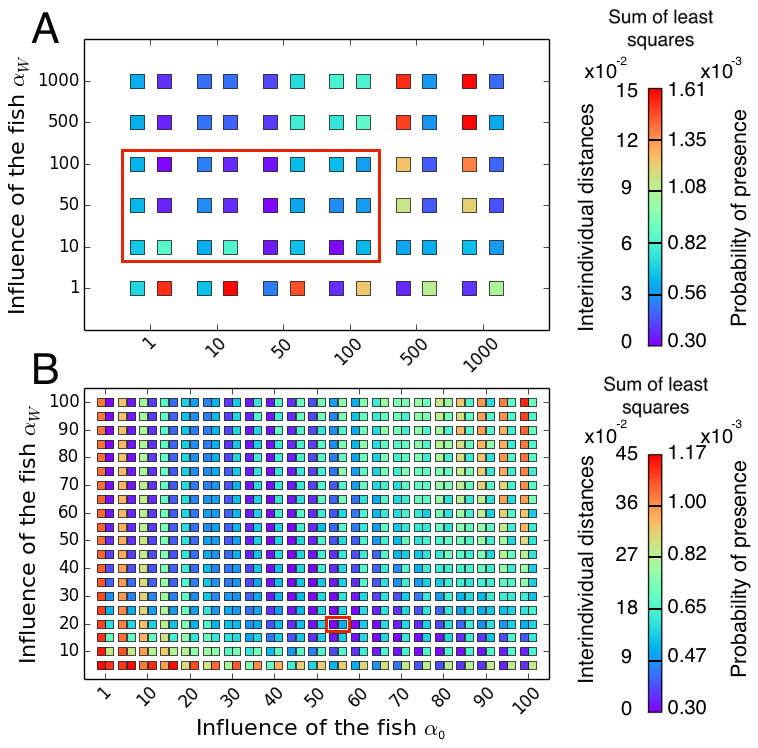}
\caption{Fitting by sum of least squares minimisation of the parameters $\alpha_0$ and $\alpha_W$ determining the interaction of a focal fish that far away or close to a wall with other individuals. To each couple ($\alpha_0$, $\alpha_W$) corresponds a couple of rectangles whose colours indicate the value of the sum of least squares obtained for the comparison of the interindividual distances (left rectangle) and the probability of presence (right rectangle) with the experimental data shown in Fig.~\ref{figure9}A and Fig.~\ref{figure11}B. Lower values of sum of least squares corresponding to better fits are in the purple and blue colour scale. (A) The exploration of a first set of parameter values indicated that the best couples ($\alpha_0$, $\alpha_W$) are found for values ($\alpha_0 \le 100$, $10 \le \alpha_W \le 100$), highlighted in the red rectangle. (B) The refinement of the parameters exploration in the ranges determined in (A) showed that the best fitting is obtained by $\alpha_0 = 55$ and $\alpha_W = 20$.}
\label{figure10}
\end{figure}

\begin{figure}[ht]
\centering
    {
      \includegraphics[width=0.5\textwidth]{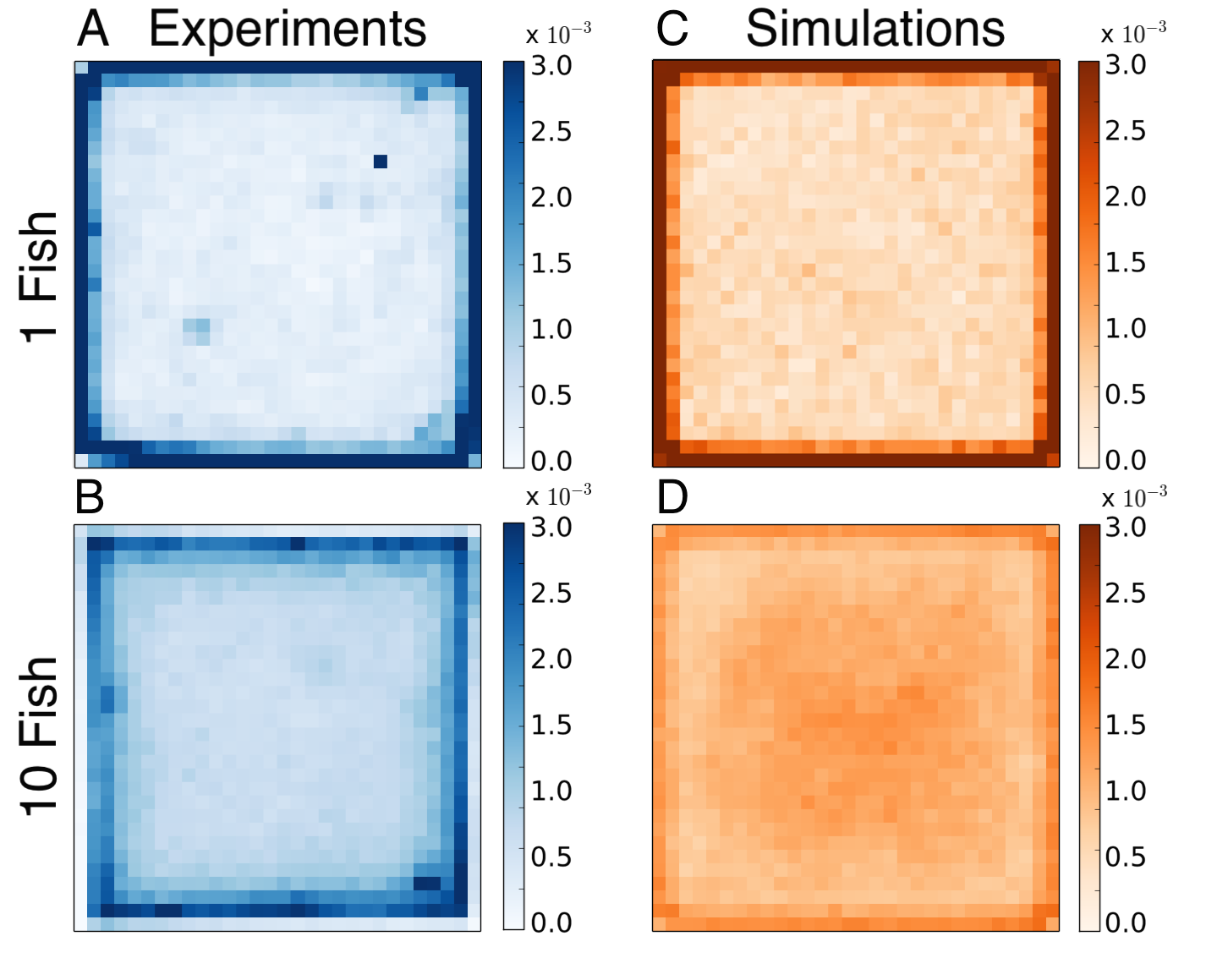}
    }
  \caption{Probability of presence for experimental (A, B) and simulated (C, D) data. Results are obtained for 10 replicates of 1h. (A) Experimental individual behaviour of single fish from AB strain in the empty experimental tank. The probability of presence showed that fish were mainly swimming along the walls of the tank. (B) Experimental results obtained for groups of 10 AB strain zebrafish in a homogeneous environment. Similarly to single fish, groups of 10 zebrafish were mainly swimming along the walls of the tank. (C) Probability of presence of a simulated single fish in a homogeneous environment during one hour (results for 10 simulations). (D) Probability of presence of a simulated group of 10 fish in a homogeneous environment during one hour (results for 10 simulations).}
  \label{figure11}  
\end{figure}

To simulate experiments with 10 fish, we introduce three parameters describing the interaction with the fish: the measure of concentration $\kappa_F$ associated with the PDF computed for each fish and the parameters $\alpha_0$ and $\alpha_W$ that weight the influence of other individuals on a fish that is far from ($\alpha_0$) or close ($\alpha_W$) to a wall. The value of $\kappa_F$ is assumed to be similar than $\kappa_W$ and is equal to 20. By doing so, we consider that a fish orients towards a given target $\mu$ with a high accuracy. To determine the value of the weights $\alpha_*$ we perform simulations with different values of ($\alpha_0$, $\alpha_W$) and compare the distribution of interindividual distances and probability of presence with those obtained for the experiments. The best values to reproduce both distributions are $\alpha_0 = 55$ and $\alpha_W = 20$ (Fig.~\ref{figure10}). Thus, fish that follow a wall are less influenced by other congeners than fish situated in the center of the tank. With these parameter values, our model was able to reproduce the probability of presence displayed by groups of 10 zebrafish as shown in Fig.~\ref{figure11}D. Concerning the distribution of interindividual distances, the model reproduces the decreasing distribution observed in the experiments except for the mode of the distribution that is between 0m and 0.1m (Fig.~\ref{figure9}B).

\subsection{Heterogeneous environment}

\begin{figure}[ht]
\centering
   {
      \includegraphics[width=0.41\textwidth]{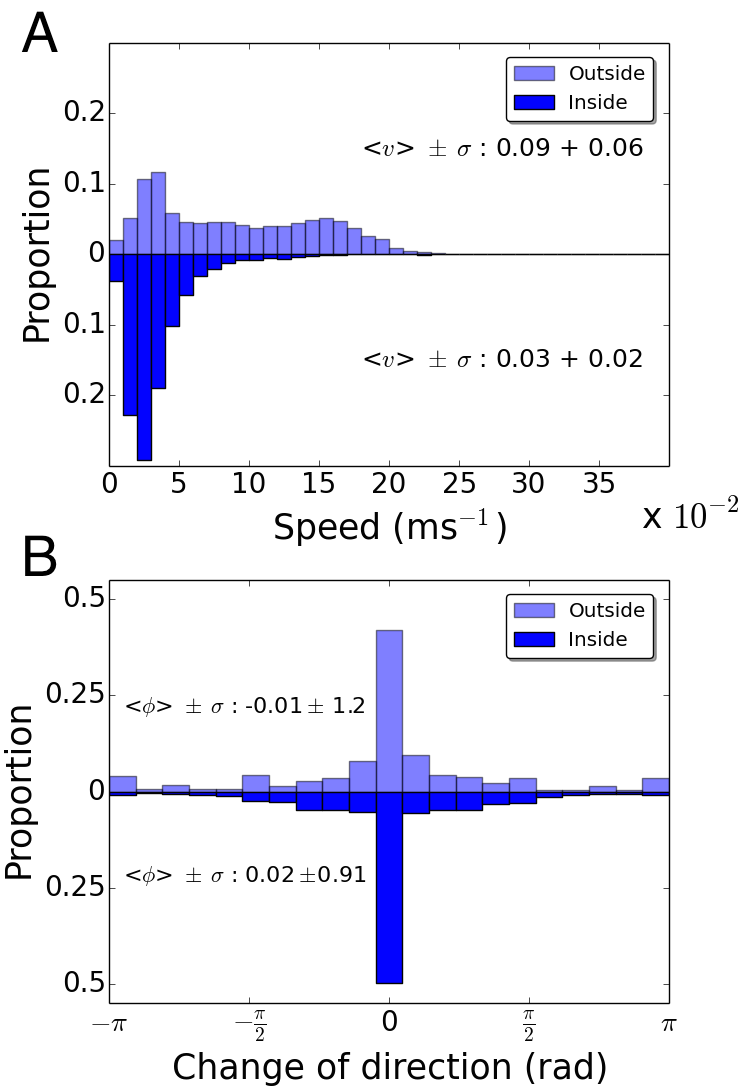}
    }
  \caption{Experimental individual behaviour of single fish from AB strain in the experimental tank with two spots of interest. (A) The distribution of the speed shows an average speed of 0.09 $\pm$ 0.06 ms$^{-1}$ outside the spots while fish were swimming with an average speed of 0.03 $\pm$ 0.02 ms$^{-1}$ under the spots. (B) The distribution of the change in orientation highlights that fish are mainly swimming forward with low deviation both outside and inside the spots. Results are cumulated for 10 replicates of one hour.
  }
  \label{figure12}  
\end{figure}

\begin{figure}[ht]
\centering
    {
      \includegraphics[width=0.5\textwidth]{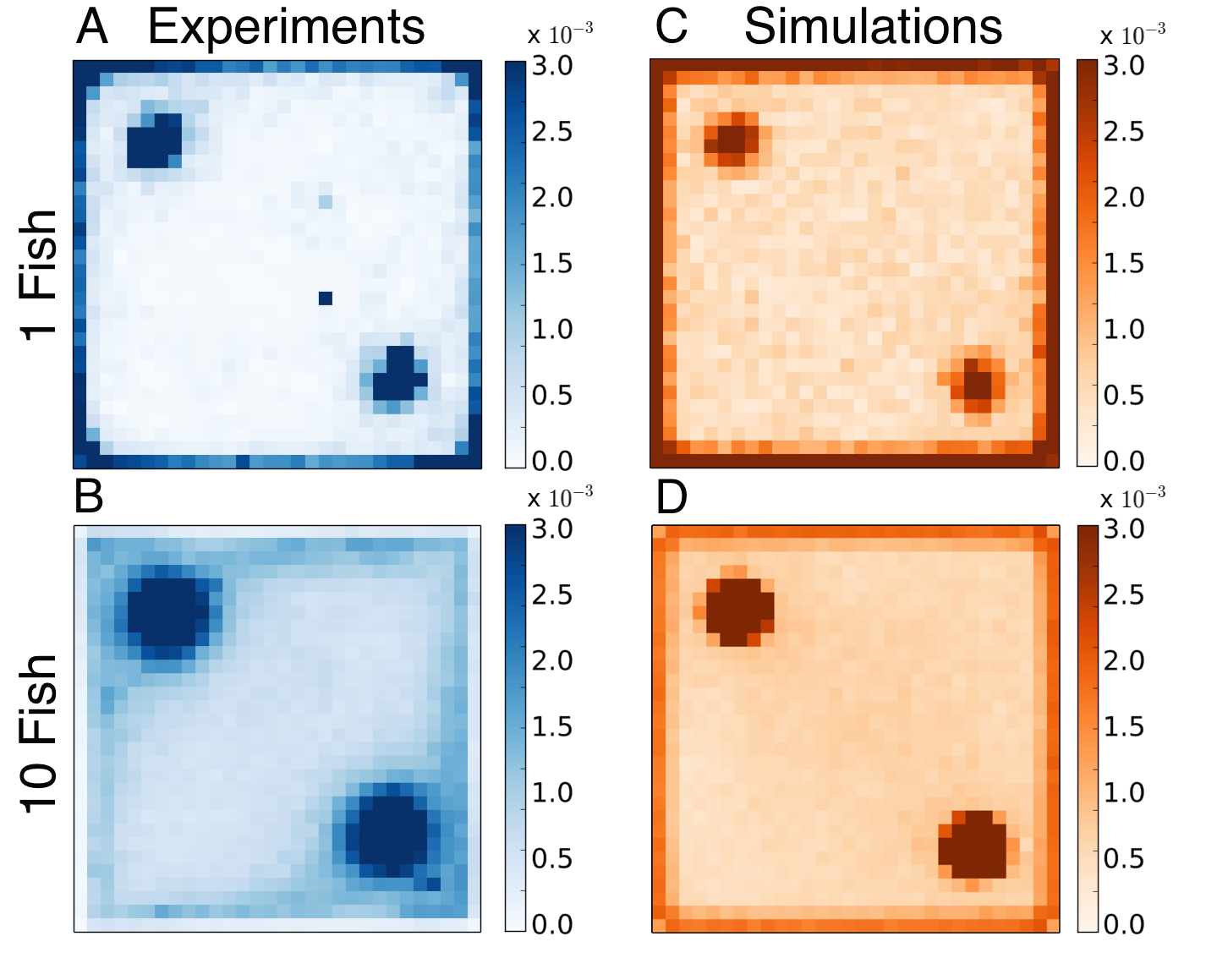}
    }
  \caption{Probability of presence for experimental (A, B) and simulated (C, D) data. Results are obtained for 10 replicates of 1h. (A) Experimental individual behaviour of a single fish from AB strain in a tank with two floating discs. The probability of presence showed that fish were mainly swimming along the walls or under the discs. (B) Experimental results obtained for groups of 10 AB strain zebrafish in a heterogeneous environment. The probability of presence showed that fish were mainly present under the floating disks. Results are obtained from 10 replicates of one hour. (C) Probability of presence of a simulated single fish in a homogeneous environment during one hour (results for 10 simulations). (D) Probability of presence of a simulated group of 10 fish in a homogeneous environment during one hour (results for 10 simulations).}
  \label{figure13}  
\end{figure}

We added two spots of interest in the experimental tank placed at 25$\sqrt{2}$ cm from two opposite corners along the diagonal of the tank. These spots consisted of blue plastic disks (20cm of $\diameter$) floating at the water surface and hung by nylon threads. An example of path followed by a fish during 10 minutes is shown on Fig.~\ref{figure5}B. We calculated similar parameters from the individual positions of fish (10 x 1 fish) moving alone in the tank with two spots. In the presence of two spots, the fish are mainly detected along the walls and under the spots as shown by their probability of presence (Fig.~\ref{figure13}A). Experiments with single fish also showed that fish decreased their speed under these spots. Indeed, the separation of the speed distribution measured \textit{Outside} or \textit{Inside} the spots shows that the average speed was three times slower when fish were located under a floating disk (Fig.~\ref{figure12}A). On the contrary, the presence of the floating disks did not affect the distribution of the changes in orientation that were similar \textit{Outside} and \textit{Inside} (Fig.~\ref{figure12}B).

\begin{figure}[h!]
\centering
\includegraphics[width=.5\textwidth]{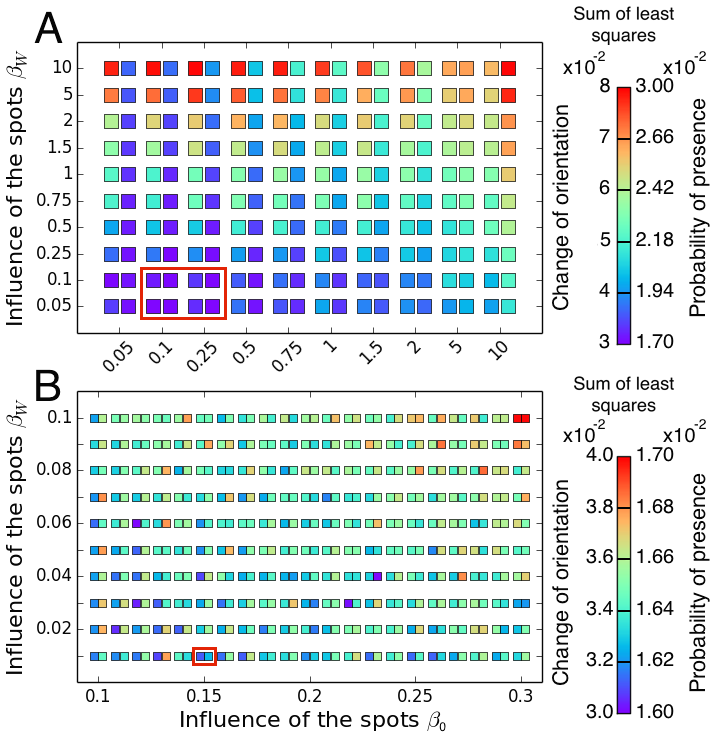}
\caption{Fitting by sum of least squares minimisation of the parameters $\beta_0$ and $\beta_W$ determining the interaction of a focal fish that far away or close to a wall with the spots of interest. To each couple ($\beta_0$, $\beta_W$) corresponds a couple of rectangles whose colours indicate the value of the sum of least squares obtained for the changes of orientation (left square) and the probability of presence (right square) with the experimental data shown in Fig.~\ref{figure12}B and Fig.~\ref{figure13}A. Lower values of sum of least squares corresponding to better fits are in the purple and blue colour scale. (A) The exploration of a first set of parameter values indicated that the best couples ($\beta_0$, $\beta_W$) are found for values ($0.05 \le \beta_0 \le 0.5$, $ 0 \le \beta_W \le 0.25$), highlighted in the red rectangle. (B) The refinement of the parameters exploration in the ranges determined in (A) showed that the best fitting is obtained by $\beta_0 = 0.15$ and $\beta_W = 0.01$.}
\label{figure14}
\end{figure}

For experiments involving a single fish, only the interaction with the walls and the shelters of the tank are present. Since $d_w = 0.05$ and $\kappa_w = 20$ were fitted by our experiments in homogeneous environment, we explore the parameter values of $\beta_0$ and $\beta_W$ (the ponderation of the influence of the spots) and assume that the measure of concentration associated with the spots $\kappa_S = \kappa_w = 20$. We perform simulations for different values of $(\beta_0, \beta_W)$ and compare their results with the experimental probability of presence and distribution of change of orientation. The best value to reproduce our experimental data are $\beta_0 = 0.15$ and $\beta_W = 0.01$ as shown by Fig.~\ref{figure14}. As observed for the influence of the congeners, these values indicate that fish following a wall are less influenced by the spots than fish swimming in the centre of the tank. With these parameter values, the model reproduce the spatial distribution of the fish along the wall and under the floating disks (Fig.~\ref{figure13}C).

Finally, we observed 10 groups of 10 zebrafish swimming in the presence of two spots of interest during one hour. In this case, the fish were also observed mainly under the spots and along the walls but show a preference for the spots (Fig.~\ref{figure13}B). The measure of the inter-individual distances shows that the presence of spots does not have a strong influence on the distance between the fish (Fig.~\ref{figure16}A). Indeed, the average distance between the individuals is $0.439$m $\pm 0.331$m, which is 6mm less than observed in the absence of floating disks.

\begin{figure}[h!]
\centering
\includegraphics[width=.5\textwidth]{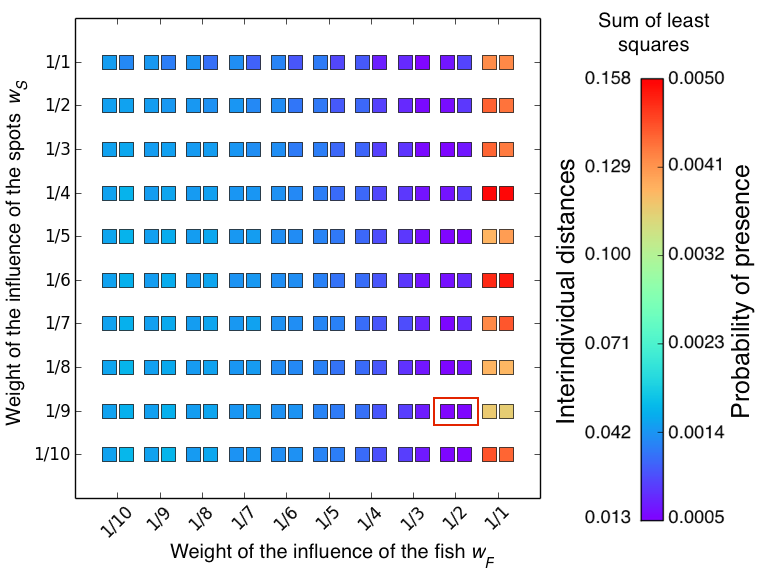}
\caption{Fitting by sum of least squares minimisation of the parameters weighting the influence of the fish $w_F$ and the spots $w_S$ in simulation with 10 zebrafish swimming with two spots of interest. To each couple ($w_F$, $w_S$) corresponds a couple of rectangles whose colours indicate the value of the sum of least squares obtained for the comparison of the interindividual distances (left square) and the probability of presence (right square) with the experimental data shown in Fig.~\ref{figure16}A and Fig.~\ref{figure13}B. Lower values of sum of least squares corresponding to better fits are in the purple and blue colour scale. The exploration of a set of parameter values indicated that the best couples ($w_F$, $w_S$) is $w_F=\frac{1}{2}$, $w_S=\frac{1}{9}$.}
\label{figure15}
\end{figure}

\begin{figure}[ht]
\centering
\includegraphics[width=.4\textwidth]{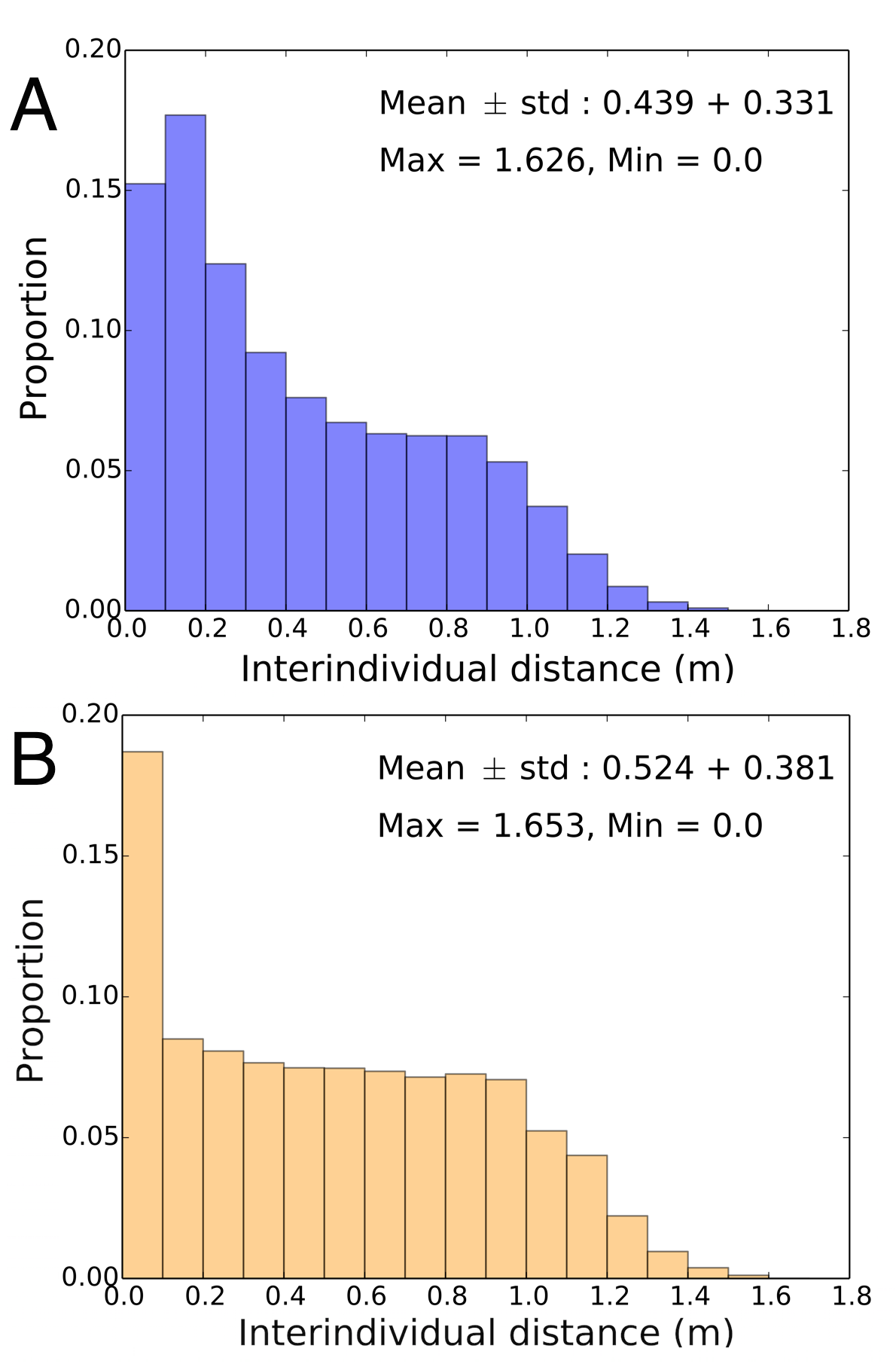}
\caption{Cumulated interindividual distances measured between all pairs of fish swimming in the presence of two spots of interest. Results are obtained for 10 experiments of one hour with 10 groups of 10 fish (A) and 10 simulations with 10 agents (B).}
\label{figure16}
\end{figure}

We simulated these experiments with groups of 10 zebrafish with our model by integrating the interactions of the agents with both the other fish and the spots of interest. In the previous experiments, the parameter ruling the interactions with the fish ($\alpha_0$, $\alpha_W$) and the spots of interest ($\beta_0$, $\beta_W$) were fitted independently but here, both stimuli are simultaneously present in the perception field of the fish. Therefore, we investigate the relative importance of both stimuli by weighting the influence of the other fish or the spots of interest. To do so, we multiply the previously fitted values of ($\alpha_0$, $\alpha_W$) and ($\beta_0$, $\beta_W$) by different weighting factors $w_F = \{\frac{1}{1}, \frac{1}{2} ... \frac{1}{10} \}$ and $w_S = \{\frac{1}{1}, \frac{1}{2} ... \frac{1}{10} \}$. For example, values $w_F=\frac{1}{1}$ and $w_S=\frac{1}{1}$ imply that the two stimuli are simply added while values $w_F=\frac{1}{2}$ and $w_S=\frac{1}{5}$ imply that the influence of the fish is divided by 2 and the influence of the spots by 5. Thus, we perform simulations for each couple of ($w_F, w_S$) and compare the distributions of the probability of presence and the interindividual distances with those measured experimentally. The best fit is given by $w_F = 2$ and $w_S=9$ (Fig.~\ref{figure15}) which implies that the influences of the fish and the spots have to be decreased by a factor 2 and 9 respectively. Therefore, when both stimuli (congeners and spots) are perceived by the fish, the best values of $\alpha_*$ and $\beta_*$ to reproduce the experimental data are $\alpha_0 = 27.5$, $\alpha_W = 10$, $\beta_0 = 0.016$ and $\beta_W = 0.0011$. While these values give a correct fit of the experimental probability of presence (Fig.~\ref{figure13}D), the model do not perfectly reproduce the distribution of the interindividual distances (Fig.~\ref{figure16}B). Similarly than for the groups of fish in a homogeneous environment, the mode interindividual distances measured in the simulation is between 0 and 0.1m while the mode of the experimental distances is between 0.1 and 0.2m. 

\section{Discussion}

\subsection{Experimental results}

In this study, we observed individual and collective behaviour of zebrafish in homogeneous and heterogeneous environments. Firstly, we observed single \textit{D. rerio} swimming in our experimental tank to describe the swimming behaviour of zebrafish. In homogeneous environment, the fish were mainly following the wall of the tank and avoid the center of it. This observation was also reported in studies performed with zebrafish and robotic-fish \cite{Butailetal.2014}. We characterized the motion patterns of single individuals and found out an average speed of $0.07 \pm 0.03$ m/s and an average change of orientation of $-0.03 \pm 0.84$ rad, in coherence with previous studies performed on single zebrafish individuals \cite{Langeetal.2013, Zienkiewiczetal.2014, Mwaffoetal.2014}. Our experiments with 10 zebrafish showed that the spatial repartition of the group did not differ significantly from the individual one. Fish in groups were also mainly detected along the walls of the tank.

The presence of floating disks influence the spatial distribution of the fish in the experimental tank. The shaded area seemed as attractive as the walls since fish showed a similar probability of presence for both stimuli. These disks had no influence on the change of direction of the fish but resulted in a spatial differentiation of the instantaneous speed of the individuals. Indeed, the fish swam with an average speed of $0.09 \pm 0.06$ m/s outside the shelters but with an average speed of $0.03 \pm 0.02$ m/s under them. Such reduction of speed can indicate that shaded areas are potentially considered as temporary resting sites by the fish. The observation of groups of 10 fish showed similar results with preference for both the walls and the shaded area.

These results show that zebrafish are avoiding free water and prefer to swim near potential shelters (floating objects or bank). Although fish were attracted by the floating disk, they did not seem to avoid the light since their presence under the disk and along the wall (that are exposed to light) are similar. These observations are in accordance with the ecology of the species that shows a diurnal activity in the Nepalese and Indian shallow water and rice paddies \cite{Parichy2015}.

\subsection{A new hybrid modeling approach}

Based on our observations and on the literature, we developed a model describing individual and collective motion. Currently, collective motion receives attention from researchers of numerous fields such as biology, physics, computer science and robotics. Current modeling tends to give up traditional methods (metric or topologic) accounting for the perception of influential neighbors by the focal fish for vision-based approach \cite{Lemassonetal.2009, Lemassonetal.2013, Strandburg-Peshkinetal.2013}. Here we present a 3D perception system accounting for zebrafish vision in which the objects are described by the solid angle that they capture in their perception field. By doing so, we are closer to a realistic description of the sensory system of the fish. In this first step, we assumed the simple hypothesis that object were homogeneously perceived in this 3D sensory field. However, the recent characterization of the visual system of zebrafish highlighted centers of acute vision (i.e. \textit{areae}) in the fronto-dorsal region \cite{Pitaetal.2015}. Therefore, future models should take into account such heterogeneity in the perception field. In addition, the position of the areae differs from one species to the other and is expected to influence the structure of the fish school \cite{Pitaetal.2015}. The extension of our model to 3 dimensional movements would allow to test such hypothesis.

In addition to this sensory system, we proposed a new mechanism to determine the direction of an agent according to its visual field. Rather than computing a resulting vectorial force that is applied to the fish, we redefined the modeling approach by describing the decision-making process of the fish to make an intentional movement according to its perception field. In this first step, we represented the choice of the individual by a stochastic process. The fish can potentially move in any direction but it will favor directions associated with a perceived stimuli (congeners for example). This stochastic model can reproduce collective behaviour exhibited by species that show cohesive behaviour without presenting higher order or only occasionally without introducing a high level of noise accounting for disturbances. This is made possible by the representation of the perception field of the fish and its translation into a probability distribution function. By doing so, we are closer to an effective description of the individual decision-making process during motion \cite{Pitaetal.2015}. Indeed, with such approach of PDFs' summation, we can include potentially all kinds of stimuli (congeners, environment, food...) that are perceived by the fish and account for a choice or consensus between potentially antagonist stimuli. Such choice between two concurrent stimuli has been evidenced in zebrafish larvae that orient towards one of two light source rather than swimming towards their bisector \cite{Burgessetal.2010}. The authors showed that two retinal pathways controlling turn movements and rapid forward swimming are responsible for phototaxis. Such advances on the understanding of information processing by neuronal pathways should be integrated in further multi-scale models to make the link between fish movement, perception of information and its processing based on biological knowledge.

Moreover, most of the models consider motion in unbounded homogenous space (torus geometry). While this hypothesis is reasonable to study the collective behaviour of animal living in pelagic water, the interactions with can not be neglected for species living in small streams. We extend our model to take into account bounded and heterogenous space getting closer to natural conditions of fish like \textit{D. rerio}. The model correctly reproduced the behaviour of single individual swimming in a bounded tank. The fish mainly follows the wall but sometimes swim in the center of the tank. The probability of presence was also correctly reproduce but the distribution of interindividual distances was biased towards short distances. This could originate from the absence of preferred distance or avoidance distance between the agents. Indeed, while the agents of the model can overlap, the fish have to respect a distance between them. Such distance was not introduce in this first parametrization of the model to limit the number of parameters. In a heterogeneous environment, the model was also able to reproduce the individual trajectories of single fish that transits between the walls and the spots of interest. As in the homogeneous environment, the probability of presence of groups of fish is also fitted, but again, the distribution of the interindividual distances is biased towards short distances. Therefore, these results indicate that the approach developed in this study can successfully describe individual and collective motions of fish but that further version could include additional biological variable and behaviour that are species dependent. For example, here the probability distribution function for a simulated fish are centered on the \textit{position} of the other fish that it perceives but the PDF could be centered on the \textit{direction} of the congeners to simulate an alignement behaviour. Rather than developing a model that exactly described the behaviour of zebrafish, our goal was to introduce a new decision-making algorithm for motion in complex environment.

In this study, we applied our decision making algorithm to an individual-based model (IBM). Recently, a kinetic model (KM) was proposed to account for the individual movement of isolated zebrafish \cite{Zienkiewiczetal.2014, Mwaffoetal.2014}. This model inspired by \cite{Gautraisetal.2009, Gautraisetal.2012} took into account a dynamic speed regulation characterizing \textit{D. rerio} motion. In our model, we did not implement a function accounting for speed modulation in order to focus on the decision-making mechanism. The instantaneous speed was drawn from the experimental distribution with a time step corresponding to the tail-beat period of zebrafish. Then, future work should investigate the impact of the proposed stochastic mechanism for orientation on such KMs. Moreover, the development of a KM version of our IBM could be an intermediate step towards a continuum model (CM) description of the group of agents. This could then be use to perform large-scale analysis and prediction of the collective behaviour displayed by very large population. Such multiscale modeling approach would allow us to identify the properties of the group that are preserved at all scales of analysis or on the contrary that are specific to a particular level of observation \cite{Degondetal.2013}. We performed simulations with a small number of agents to mimic our experimental conditions in this study that are close to the natural size of zebrafish group \cite{Parichy2015}. While our IBM could model larger groups of tens of individuals, simulations involving thousands of agents would require a longer computational time than KM and CM.

In parallel to the understanding of information processing by individuals, this approach is also a new step towards bio-inspired algorithms that can be implemented in robotic agents. Indeed, it is a major scientific challenge to build artificial systems composed of robots that can perceive, communicate to, interact with other agents (biological or robotic) and adapt to their environment \cite{Schmickletal.2013}. To do so, we need to develop artificial agents that communicate through appropriate channels corresponding to specific animal traits but also that correctly perceive and interpret signals emitted by the animals \cite{Halloyetal.2013, Mondadaetal.2013}. This was firstly achieved in 2007 by building bio-inspired artificial cockroaches that where able to sense the presence of congeners and to adapt their behaviour following a bio-inspired algorithm \cite{Halloyetal.2007}. 

In fish, an increasing number of studies aim at developing such robotic agents to interact with group of fish \cite{Fariaetal.2010, Abaidetal.2012, Landgrafetal.2013, Swainetal.2012}. Current experiments generally involve one robot that follows a predetermined trajectory or that moves according to the position of fish detected through the intermediary of a camera that films the entire setup. While this methodology is in the line of the classical ethological experimentation to investigate behavioural responses of animals, the development of fully autonomous integrated lures in fish schools (or other species) requires the development of perception abilities for the robotic agents as well as adapted behavioural algorithms. 

In this perspective, the development of perception-based models is a necessary step towards intelligent artificial systems capable of closing the loop of interaction between animals and robots.


\section{Methods}

\subsection{Animals and housing}

We acquired 150 adult wild-type zebrafish (\textit{Danio rerio} AB strain from Institut Curie (Paris). Fish were 18 months old at the time of the experiments. We kept fish under laboratory conditions, $27\,^{\circ}{\rm C}$, 500$\mu$S salinity with a 10:14 day:night light cycle. The fish were reared in 55 litres tanks and were fed two times per day (Special Diets Services SDS-400 Scientific Fish Food). Water pH was maintained at 7.5 and Nitrites (NO$^{2-}$) are below 0.3 mg/l.

\subsection{Experimental setup}

We recorded the behaviour of zebrafish in a 120 x 120 x 30 cm experimental tank made of glass with internal walls covered with white adhesive. The water depth was kept at 10cm during the experiments. A Logitech\textregistered HD Pro Webcam C920 was mounted 160 cm above the water surface to record experiment at a resolution 1920 x 1080 and at 15 FPS. The camera was connected to a workstation Dell\textregistered Precision T5600 dedicated to the recording of the videos and their analysis. One halogen lamp (450W) was placed at each corner of the tank and oriented towards the wall to provide indirect lightning of the tank. The whole set-up was confined behind white sheets to isolate experiments and homogenize luminosity.

\subsection{Experimental procedure}

We recorded the behaviour of zebrafish swimming in our tank during one hour. We tested two numbers of individuals (single fish or groups of 10 fish) in two environmental conditions (homogeneous or heterogeneous). The heterogeneous environment was created by adding two floating disks made of blue coloured Plexiglas (200 mm diameter and 3 mm thick) suspended by nylon threads. The two spots were spaced from 70 cm and located on a diagonal of the square. Before the trials, fish were placed with a hand net in a cylindrical arena (20 cm diameter) made of Plexiglas placed in the centre of our experimental tank. Following a five minutes acclimatization period, this arena was removed and fish were able to freely swim in the experimental tank. We performed 10 replicates for each combinaison of parameters (number of fish x environmental condition).

\subsection{Data analysis}

The recorded videos were analyzed offline using a custom Matlab script developed to detect the position of the fish. This script performed a background subtraction on each frame and transformed it in a binary image according to a pixel intensity threshold given by the user. The software then identified the blobs on this image and kept only the blobs that were formed by more than 20 and less than 200 pixels (those values were obtained by manually identifying the fish on multiple frames). Since this method did not allow a perfect detection of all the individuals, we developed a second script that was run after the first one and that plotted the frame where a fish (or more) was undetected for the user to manually identify the missing individual(s). While this analysis tool is time-costly, it allowed us to identify the fish that were partially hidden during a collision/superposition with another fish or the fish that were situated under the floating disks since our program was not able to detect them due to a lack of intensity of the pixel after the background subtraction. 
The positions $P(x,y)$ of the fish were recorded at each time step $t = 1/15s$ during the experiment with single fish in homogeneous environment and $t = 1s$ for all other experimental conditions. This allowed us to build the trajectory of each individual for experiments involving a single fish and to compute the speed of the individuals as well as their change in orientation between successive positions. The instantaneous speed $v_t$ was calculated on three positions and thus computed as the distance between $P(x,y,t-1)$ and $P(x,y,t+1)$ divided by 2 time steps while change in orientation were computed as the angle between two successive speed vectors). The distributions of speed were computed only for parts of the trajectory during which the fish were not in freezing behaviour (i.e. immobile). This corresponds to a spontaneous speed higher than 1mm per second. Since our tracking system did not resolve collision with accuracy, we did not calculate individual measures for data obtained with groups of fish but characterized the aggregation level of the group.

\subsection{Implementation and numerical simulations}

The model was implemented in a Matlab script. The simulations were run during 10800 time steps, each time step representing an increment of 1/3 second to simulate a total time of 1 hour, similarly to our experiments. This time step was chosen according to the tail beat frequency of the zebrafish of ~2.5 Hz. By doing so, we assume that zebrafish can potentially change their orientation at each tail beat.
The position of the agents on the 2D plane space is described by the position of their head (x, y, 0). The positions of the other vertices are computed according to the position of the head and the direction of the fish. To compute the solid angle of the perceived stimuli in the perception field of the focal fish, we calculated the area of the spherical polygon delimited by the projection of the stimuli's vertices on a unit sphere centered on the focal fish. To do so, we divided the polygon in two spherical triangle and computed their spherical excess using L'Huilier's theorem :

\small
\begin{equation}
\tan(\frac{1}{4}E)= \sqrt{ \tan(\frac{1}{2}s) \tan[\frac{1}{2}(s-a)] \tan[\frac{1}{2}(s-b)] \tan[\frac{1}{2}(s-c)] } 
\end{equation}
\normalsize
with $a$, $b$, and $c$ the length of the arcs between the vertices expressed in spherical coordinates ($\rho$, $\theta$, $\phi$) and computed by : 
\small
\begin{equation}
a = \arccos ( \sin(\phi_1) \sin(\phi_2) + \cos (\theta_1 - \theta_2) \cos (\theta_1) \cos (\theta_2) 
\end{equation}
\normalsize
and $s$ the semi parameter given by :

\begin{equation}
s = \frac{a + b + c }{2}
\end{equation}

\phantomsection
\section*{Acknowledgments}

The authors thank Filippo Del Bene (Institut Curie, Paris, France) that provided us the fish observed in the experiments reported in this paper. This work was supported by European Union Information and Communication Technologies project ASSISIbf, (Fp7-ICT-FET n. 601074). The funders had no role in study design, data collection and analysis, decision to publish, or preparation of the manuscript.


\addcontentsline{toc}{section}{Acknowledgments}

\begin{thebibliography}{99}

\bibitem{FriedlAndGilmour2009}
P.~Friedl and D.~Gilmour.
\newblock Collective cell migration in morphogenesis, regeneration and cancer.
\newblock {\em Nature Reviews}, 10:445--457, 2009.

\bibitem{Friedletal.2004}
P.~Friedl, Y.~Hegerfeldt, and Tusch M.
\newblock Collective cell migration in morphogenesis and cancer.
\newblock {\em International Journal of Developmental Biology}, 48:441--449,
  2004.

\bibitem{EtienneManneville2014}
S.~Etienne-Manneville.
\newblock Neighborly relations during collective migration.
\newblock {\em Current Opinion in Cell Biology}, 30:51--59, 2014.

\bibitem{Sokolovetal.2009}
A.~Sokolov, R.E. Goldstein, F.I. Feldchtein, and I.S. Aranson.
\newblock Enhanced mixing and spatial instability in concentrated bacterial
  suspensions.
\newblock {\em Physical Review E}, 80:031903, 2009.

\bibitem{BanJacob2008}
E.~Ben-Jacob.
\newblock Social behavior of bacteria: from physics to complex organization.
\newblock {\em European Physical Journal B}, 65:315--322, 2008.

\bibitem{Wolgemuth2008}
C.W. Wolgemuth.
\newblock Collective swimming and the dynamics of bacterial turbulence.
\newblock {\em Biophysical Journal}, 95:1564--1574, 2008.

\bibitem{Zhangetal.2010}
H.P. Zhang, A.~Be'er, E.L. Florin, and H.L. Swinney.
\newblock Collective motin and density fluctuations in bacterial colonies.
\newblock {\em Proceedings of the National Academy of Science USA},
  107:13626--13630, 2010.

\bibitem{Bazazietal.2008}
S.~Bazazi, J.~Buhl, J.J. Hale, M.L. Anstey, G.A. Sword, S.J. Simpson, and I.D.
  Couzin.
\newblock Collective motion and cannibalism in locust migratory bands.
\newblock {\em Current Biology}, 18:735--739, 2008.

\bibitem{Buhletal.2006}
J.~Buhl, D.J.T. Sumpter, I.D. Couzin, J.J. Hale, E.~Despland, E.R. Miller, and
  S.J. Simpson.
\newblock From disorder to order in marching locusts.
\newblock {\em Science}, 312:1402--1406, 2006.

\bibitem{Simpsonetal.2006}
S.J. Simpson, G.A. Sword, P.D. Lorch, and I.D. Couzin.
\newblock Cannibal crickets on a forced march for protein and salt.
\newblock {\em Proceedings of the National Academy of Science USA},
  103:4152--4156, 2006.

\bibitem{Deneubourgetal.1989}
J.L. Deneubourg, S.~Goss, N.R. Franks, and J.M. Pasteels.
\newblock The blind leading the blind: modeling chemically mediated army ant
  raid patterns.
\newblock {\em Journal of Insect Behavior}, 2:719--725, 1989.

\bibitem{CouzinAndFranks2003}
I.D. Couzin and N.R. Franks.
\newblock Self-organized lane formation and optimize traffic flow in army ants.
\newblock {\em Proceedings of the Royal Society of London B}, 270:139--146,
  2003.

\bibitem{Jansonetal.2005}
S.~Janson, M.~Middendorf, and M.~Beekman.
\newblock Honeybee swarms: how do scouts guide a swarm of uninformed bees.
\newblock {\em Animal Behaviour}, 70:349--358, 2005.

\bibitem{Ballerinietal.2008}
M.~Ballerini, N.~Cabibbo, R.~Candelier, A.~Cavagna, E.~Cisbani, I.~Giardina,
  A.~Orlandi, G.~Parisi, A.~Procaccini, M.~Viale, and V.~Zdravkovic.
\newblock Empirical investigation of starling flocks: a benchmark study in
  collective animal behaviour.
\newblock {\em Animal Behaviour}, 76:201--215, 2008.

\bibitem{LebarBajecAndHeppner2009}
I.~Lebar~Bajec and F.H. Heppner.
\newblock Organized flight in birds.
\newblock {\em Animal Behaviour}, 78:777--789, 2009.

\bibitem{KingAndSumpter2012}
A.J. King and D.J.T. Sumpter.
\newblock Murmurations.
\newblock {\em Current Biology}, 22:R112--R114, 2012.

\bibitem{HemelrijkAndKunz2005}
C.K. Hemelrijk and H.~Kunz.
\newblock Density distribution and size sorting in fish schools: an
  individual-based model.
\newblock {\em Behavioral Ecology}, 16:178--187, 2005.

\bibitem{Parrishetal.2002}
J.K. Parrish, S.V. Viscido, and D.~Gr{\"u}nbaum.
\newblock Self-organized fish schools: an examination of emergent properties.
\newblock {\em Biological Bulletin}, 202:296--305, 2002.

\bibitem{Beccoetal.2006}
C.~Becco, N.~Vandewalle, J.~Delcourt, and P.~Poncin.
\newblock Experimental evidences of a structural and dynamical transition in
  fish school.
\newblock {\em Physica A}, 267:487--493, 2006.

\bibitem{Fischhoffetal.2007}
I.R. Fischhoff, S.R. Sundaresan, J.~Cordingley, H.M. Larkin, M.J. Sellier, and
  D.I. Rubenstein.
\newblock Social relationships and reproductive state influence leadership
  roles in movements of plains zebra, \textit{Equus burchellii}.
\newblock {\em Animal Behaviour}, 73:825--831, 2007.

\bibitem{Helbingetal.2001}
D.~Helbing, P.~Molnar, I.J. Farkas, and K.~Bolays.
\newblock Self-organizing pedestrian movement.
\newblock {\em Environment and Planning B: Planning and Desgin}, 28:361--383,
  2001.

\bibitem{Moussaidetal.2011}
M.~Moussa{\"\i}d, D.~Helbing, and G.~Theraulaz.
\newblock How simple rules determine pedestrian behavior and crowd disasters.
\newblock {\em Proceedings of the National Academy of Science USA},
  108:6884--6888, 2011.

\bibitem{Aoki1982}
I.~Aoki.
\newblock A simulation study on the schooling mechanism in fish.
\newblock {\em Bulletin of the Japanese Society for the Science of Fish},
  48:1081--1088, 1982.

\bibitem{Aoki1984}
I.~Aoki.
\newblock Internal dynamics of fish schools in relation to inter-fish distance.
\newblock {\em Bulletin of the Japanese Society for the Science of Fish},
  50:751--758, 1984.

\bibitem{Bodeetal.2010}
N.W.F. Bode, J.J. Faria, D.W. Franks, J.~Krause, and A.J. Wood.
\newblock How perceived threat increases synchronization in collectively moving
  animal groups.
\newblock {\em Proceedings of the Royal Society of London B}, 277:3065--3070,
  2010.

\bibitem{Bodeetal.2011}
N.W.F. Bode, D.W. Franks, and A.J. Wood.
\newblock Limited interactions in flocks: relating model simulations to
  empirical data.
\newblock {\em Journal of the Royal Society Interface}, 8:301--304, 2011.

\bibitem{Reynolds1987}
C.W. Reynolds.
\newblock Flocks, herds, and schools: a distributed behavioral model.
\newblock {\em Computer Graphics}, 21:25--34, 1987.

\bibitem{Couzinetal.2002}
I.D. Couzin, J.~Krause, R.~James, G.D. Ruxton, and N.R. Franks.
\newblock Collective memory and spatial sorting in animal groups.
\newblock {\em Journal of Theoretical Biology}, 218:1--11, 2002.

\bibitem{Lopezetal.2012}
U.~Lopez, J.~Gautrais, I.D. Couzin, and G.~Theraulaz.
\newblock From behavioural analyses to models of collective motion in fish
  schools.
\newblock {\em Interface Focus}, 2:693--707, 2012.

\bibitem{Niwa1994}
H.S. Niwa.
\newblock Self-organizing dynamic model of fish schooling.
\newblock {\em Journal of Theoretical Biology}, 171:123--136, 1994.

\bibitem{Niwa1996}
H.S. Niwa.
\newblock Newtonian dynamical approach to fish schooling.
\newblock {\em Journal of Theoretical Biology}, 181:47--63, 1996.

\bibitem{Vicseketal.1995}
T.~Vicsek, A.~Czirok, E.~Ben-Jacob, I.~Cohen, and O.~Shochet.
\newblock Novel type of phase transition in a system of self-driven particles.
\newblock {\em Physical Review Letters}, 75:1226--1229, 1995.

\bibitem{Bertinetal.2006}
E.~Bertin, M.~Droz, and G.~Gr{\'e}goire.
\newblock Boltzmann and hydrodynamic description for self-propelled particles.
\newblock {\em Physical Review E}, 74:022101, 2006.

\bibitem{Chateetal.2008}
H.~Chat{\'e}, F.~Ginelli, G.~Gr{\'e}goire, and F.~Raynaud.
\newblock Collective motion of self-propelled particles interacting without
  cohesion.
\newblock {\em Physicel Review E}, 77:046113, 2008.

\bibitem{Chateetal.2010}
H.~Chat{\'e}, F.~Ginelli, G.~Gr{\'e}goire, F.~Peruani, and F.~Raynaud.
\newblock Modeling collective motion: variations on the vicsek model.
\newblock {\em The European Physical Journal B}, 64:451--456, 2008.

\bibitem{Nagaietal.2015}
K.H. Nagai, Y.~Sumino, R.~Montagne, I.S. Aranson, and H.~Chat{\'e}.
\newblock Collective motion of self-propelled particles with memory.
\newblock {\em Physical Review Letters}, 114:168001, 2015.

\bibitem{Gautraisetal.2009}
J.~Gautrais, C.~Jost, M.~Soria, S.~Campo, A. ans~Motsch, R.~Fournier,
  S.~Blanco, and G.~Theraulaz.
\newblock Analysing fish movement as a persistent turning walker.
\newblock {\em Journal of Mathematical Biology}, 58:429--445, 2009.

\bibitem{Gautraisetal.2012}
J.~Gautrais, F.~Ginelli, R.~Fournier, S.~Blanco, M.~Soria, H.~Chat{\'e}, and
  G.~Theraulaz.
\newblock Deciphering interactions in moving animal groups.
\newblock {\em PLoS Computational Biology}, page e1002678, 2012.

\bibitem{Zienkiewiczetal.2014}
A.~Zienkiewicz, D.A.W. Barton, M.~Porfiri, and M.~di~Bernardo.
\newblock Data-driven stochastic modelling of zebrafish locomotion.
\newblock {\em Journal of Mathematical Biology}, 2014.

\bibitem{Mwaffoetal.2014}
V.~Mwaffo, R.P. Anderson, S.~Butail, and M.~Porfiri.
\newblock A jump persistent turning walker to model zebrafish locomotion.
\newblock {\em Journal of the Royal Society Interface}, 12:20140884, 2014.

\bibitem{Moralesetal.2004}
J.M. Morales, D.T. Haydon, J.~Frair, K.E. Holsinger, and J.M. Fryxell.
\newblock Extracting more out of relocation data: building movement models as
  mixtures of random walks.
\newblock {\em Ecology}, 85:2436--2445, 2004.

\bibitem{Raichlenetal.2014}
D.A. Raichlen, B.M. Wood, A.D. Gordon, A.Z.P. Mabulla, F.W. Marlowe, and
  H.~Pontzer.
\newblock Evidence of l{\'e}vy walk foraging patterns in human
  hunter-gatherers.
\newblock {\em Proceedings of the National Academy of Science USA},
  111:728--733, 2014.

\bibitem{Lemassonetal.2009}
B.H. Lemasson, J.J. Anderson, and R.A. Goodwin.
\newblock Collective motion in animal groups from a neurobiological
  perspective: the adapive benefits of dynamic sensory loads and selection
  attention.
\newblock {\em Journal of Theoretical Biology}, 261:501--510, 2009.

\bibitem{Lemassonetal.2013}
B.H. Lemasson, J.J. Anderson, and R.A. Goodwin.
\newblock Motion-guided attention promotes adaptive communications during
  social navigation.
\newblock {\em Proceedings of the Royal Society of London B}, 280:20122003,
  2013.

\bibitem{Strandburg-Peshkinetal.2013}
A.~Strandburg-Peshkin, C.R. Twomey, N.W.F. Bode, A.B. Kao, Y.~Katz, C.C
  Ioannou, S.B. Rosenthal, C.J. Torney, H.S. Wu, S.A. Levin, and I.D. Couzin.
\newblock Visual sensory networks and effective information transfer in animal
  groups.
\newblock {\em Current Biology}, 23:R709, 2013.

\bibitem{Pitaetal.2015}
D.~Pita, B.A. Moore, L.P. Tyrrell, and Fernandez-Juricic.
\newblock Vision in two cyprinid fish: implications for collective behavior.
\newblock {\em PeerJ}, 3(e1113), 2015.

\bibitem{Burgessetal.2010}
H.A. Burgess, H.~Schoch, and M.~Granato.
\newblock Distinct retinal pathways drive spatial orientation behaviors in
  zebrafish navigation.
\newblock {\em Current Biology}, 20:381--386, 2010.

\bibitem{Butailetal.2014}
S.~Butail, G.~Polverino, P.~Phamduy, F.~Del~Sette, and M.~Porfiri.
\newblock Influence of robotic shoal size, configuration, and activity on
  zebrafish behavior in a free-swimming environment.
\newblock {\em Behavioural Brain Research}, 275:269--280, 2014.

\bibitem{Langeetal.2013}
M.~Lange, F.~Neuzeret, B.~Fabreges, C.~Froc, S.~Bedu, L.~Bally-Cuif, and W.H.J.
  Norton.
\newblock Inter-individual and inter-strain variations in zebrafish locomotory
  ontogeny.
\newblock {\em PLoS one}, 8:e70172, 2013.

\bibitem{Parichy2015}
D.M. Parichy.
\newblock Advancing biology through a deeper understanding of zebrafish ecology
  and evolution.
\newblock {\em eLife}, 4:e05635, 2015.

\bibitem{Degondetal.2013}
P.~Degond, C.~Appert-Rolland, M.~Moussa{\"\i}d, J.~Pettr{\'e}, and
  G.~Theraulaz.
\newblock A hierarchy of heuristic-based models of crowd dynamics.
\newblock {\em Journal of statistical physics}, 152:1033--1068, 2013.

\bibitem{Schmickletal.2013}
T.~Schmickl, S.~Bogdan, L.~Correia, S.~Kernback, F.~Mondada, M.~Bodi,
  A.~Gribovskiy, S.~Hahshold, D.~Miklic, M.~Szoped, R.~Thenius, and J.~Halloy.
\newblock Assisi: mixing animals with robots in a hybrid society.
\newblock In {\em Biomimetic and Biohybrid Systems Lecture Notes in Computer
  Science}, pages 441--443, 2013.

\bibitem{Halloyetal.2013}
J.~Halloy, F.~Mondada, S.~Kernback, and T.~Schmickl.
\newblock Towards bio-hybrid systems made of social animals and robots.
\newblock In {\em Biomimetic and Biohybrid Systems Lecture Notes in Computer
  Science}, pages 384--386, 2013.

\bibitem{Mondadaetal.2013}
F.~Mondada, A.~Martinoli, N.~Correll, A.~Gribovskiy, and J.~Halloy.
\newblock {\em Handbook of collective robotics: fundamentals and challenges},
  chapter A general methodology for the control of mixed natural-artificial
  societies, pages 399--428.
\newblock Pan Stanford Publishing, 2013.

\bibitem{Halloyetal.2007}
J.~Halloy, G.~Sempo, G.~Caprari, C.~Rivault, M.~Asadpour, F.~T{\^a}che,
  I.~Sa{\"\i}d, V.~Durier, S.~Canonge, J.M. Am{\'e}, C.~Detrain, N.~Correll,
  A.~Martinoli, F.~Mondada, R.~Siegwart, and J.L. Deneubourg.
\newblock Social integration of robots into groups of cockroaches to control
  self-organized choices.
\newblock {\em Science}, 318:1155--1158, 2007.

\bibitem{Fariaetal.2010}
J.J. Faria, J.R.G. Dyer, R.O. Cl{\'e}ment, I.D. Couzin, N.~Holt, A.J.W. Ward,
  D.~Waters, and J.~Krause.
\newblock A novel method for investigating the collective behaviour of fish:
  introducing 'robofish'.
\newblock {\em Behavioural Ecology and Sociobiology}, 64:1211--1218, 2010.

\bibitem{Abaidetal.2012}
N.~Abaid, T.~Bartolini, S.~Macri, and M.~Porfiri.
\newblock Zebrafish responds differentially to a robotic fish of varying aspect
  ratio, tail beat frequency, noise, and color.
\newblock {\em Behavioural Brain Research}, 233:545--553, 2012.

\bibitem{Landgrafetal.2013}
T.~Landgraf, H.~Nguyen, S.~Forgo, J.~Schneider, J.~Schr{\"o}er, C.~Kr{\"u}ger,
  H.~Matzke, R.O. Cl{\'e}ment, J.~Krause, and R.~Rojas.
\newblock Interactive robotic fish for the analysis of swarm behavior.
\newblock In {\em Advances in swarm intelligense, Lectures Notes in Computer
  Science}, pages 1--10, 2013.

\bibitem{Swainetal.2012}
D.T. Swain, I.D. Couzin, and N.E. Leonard.
\newblock Real-time feedback-controlled robotic fish for behavioral experiments
  with fish schools.
\newblock {\em Proceedings of the IEEE}, 100:150--163, 2012.

\end{thebibliography}
\end{document}